\documentclass[%
reprint,
superscriptaddress,
amsmath,
amssymb,
aps,
rmp,
floatfix,
]{revtex4-2}

\usepackage{graphicx}
\usepackage{dcolumn}
\usepackage{bm}
\usepackage{hyperref}
\usepackage[mathlines]{lineno}

\usepackage{stmaryrd} 
\usepackage{xcolor, soul}
\usepackage{placeins} 
\usepackage{subcaption} 
\usepackage{enumerate} 
\usepackage{yhmath} 

\DeclareMathOperator{\diag}{diag}

\newcommand{\E}[1]{\mathbb{E}\left[ #1 \right]}

\begin{document}

\preprint{APS/123-QED}

\title{How does liquidity shape the yield curve?}

\author{Victor Le Coz}
\email{victor.lecoz@gmail.com}
\affiliation{Chair of Econophysics and Complex Systems, \'Ecole polytechnique, 91128 Palaiseau Cedex, France}
\affiliation{LadHyX UMR CNRS 7646, \'Ecole polytechnique, 91128 Palaiseau Cedex, France}
\affiliation{Laboratoire de Math\'ematiques et Informatique pour la Complexit\'e et les Syst\`emes, CentraleSupélec, Universit\'e Paris-Saclay, 91192 Gif-sur-Yvette Cedex, France}
\affiliation{Quant AI Lab, 29 Rue de Choiseul 75002 Paris, France}

\author{Iacopo Mastromatteo}
\email{iacopo.mastromatteo@cfm.com}
\affiliation{Capital Fund Management, 23 Rue de l’Universit\'e, 75007 Paris, France}

\author{Michael Benzaquen}
\email{michael.benzaquen@polytechnique.edu}
\affiliation{Chair of Econophysics and Complex Systems, \'Ecole polytechnique, 91128 Palaiseau Cedex, France}
\affiliation{LadHyX UMR CNRS 7646, \'Ecole polytechnique, 91128 Palaiseau Cedex, France}
\affiliation{Capital Fund Management, 23 Rue de l’Universit\'e, 75007 Paris, France}

\date{\today}

\begin{abstract}
The phenomenology of the forward rate curve (FRC) can be accurately understood by the fluctuations of a stiff elastic string \citep{LeCozBouchaud-2024a}. By relating the exogenous shocks driving such fluctuations to the surprises in the order flows, we elevate the model from purely describing price variations to a microstructural model that incorporates the joint dynamics of prices and order flows, accounting for both impact and cross-impact effects. Remarkably, this framework allows for at least the same explanatory power as existing cross-impact models, while using significantly fewer parameters. In addition, our model generates liquidity-dependent correlations between the forward rate of one tenor and the order flow of another, consistent with recent empirical findings. We show that the model also account for the non-martingale behavior of prices at short timescales. 
\end{abstract}


\maketitle



\section{Introduction} \label{Introduction}

\subsection{Motivation}

The forward interest rate $f(t,T)$, which will be defined more precisely further down, represents the interest rate agreed at time $t$ for an instantaneous loan spanning from $T \geq t$ to $T+dT$. The collection of forward rates can be thought of as a continuous string that changes shape over time. An accurate understanding of the behavior of the forward interest rate curve (FRC) is essential in various fields, including interest rate derivative pricing and risk management \citep{Hull-2018, BrigoMercurio-2006}.

In an earlier article, J-P Bouchaud and one of the authors (VLC) revisited a model of the FRC based on the fluctuations of a stiff elastic string (henceforth called the BBDL model for \emph{Baaquie-Bouchaud Discrete Logarithm} model~\citep{LeCozBouchaud-2024a,BaaquieBouchaud-2004}). Compared to previous work, this approach accounts for two important features: (a) the discrete set of traded maturities, and (b) the scale-dependent structure of the correlation matrix across maturities \citep{Epps-1979}.

The objective of this article is to demonstrate that this model can be given a microstructural interpretation, which allows for new predictions. Specifically, we establish a connection between a non-measurable auxiliary noise field that appears in the construction of the original model and the physically measurable volumes traded across the interest rates curve, thus promoting the string model~\citep{LeCozBouchaud-2024a} to a microstructural model capable of predicting the price reaction to traded volumes along the curve. The resulting model is more parsimonious than other cross-impact models while maintaining comparable, if not superior, performance. We will show that it faithfully accounts for the effect of liquidity on the price-volume correlations between the forward rates of different maturities and the order flow \citep{LeCozEtAl-2024a}. Additionally, within this framework, prices appear to exhibit short-term temporal autocorrelations, consistent with established findings in the literature.

\subsection{Literature review}
\subsubsection{Arbitrage-free and field theories}
Modeling of the forward interest rate curve has been predominantly influenced by the Heath-Jarrow-Morton framework since the 1990s \citep{HeathEtAl-1992}. This seminal theory posits a finite number of risk factors, which theoretically implies the existence of risk-free portfolio directions. Addressing this limitation, various researchers have ventured beyond the conventional boundary of a finite number of driving Brownian motions \citep{Kennedy-1994,Kennedy-1997, Goldstein-2000, Santa-ClaraSornette-2001, Baaquie-2001,Baaquie-2002,Baaquie-2004, BaaquieBouchaud-2004, Cont-2005a}. In the following years, these random field theories have been applied to solve interest rate derivative pricing problems \citep{Bueno-GuerreroEtAl-2015,Bueno-GuerreroEtAl-2016,Bueno-GuerreroEtAl-2020,Bueno-GuerreroEtAl-2022, Baaquie-2007,Baaquie-2009, Baaquie-2010, Baaquie-2018, BaaquieTang-2012, BaaquieLiang-2007, WuXu-2014}.

Elaborating on the work of \citep{Baaquie-2001,Baaquie-2002,Baaquie-2004,BaaquieBouchaud-2004}, \citet{LeCozBouchaud-2024a} have proposed a random random field theory formulated on a discrete space of maturities. As previously mentioned, this approach closely reproduces the phenomenology of the FRC \citep{BouchaudEtAl-1999} and complies with the empirical finding of negligible correlation at small time scales \citep{Epps-1979}.

\subsubsection{Market micro-structure} \label{Market micro-structure}
Standard economic theory posits that an asset's price should reflect all publicly available information about its fundamental value. In reality, price formation occurs through a trading process in which information is gradually integrated into prices via the order flow of market participants. This widely recognized process is referred to as \textit{price impact}. \citet{Kyle-1985} introduced an early model of price impact, assuming a linear relationship between absolute price differences and signed volumes traded. To reconcile the temporal autocorrelation of trades \citep{BouchaudEtAl-2018, BouchaudEtAl-2004,LilloFarmer-2004,BouchaudEtAl-2009, YamamotoLeBaron-2010,TothEtAl-2015} with the temporal independence of price increments, \citet{BouchaudEtAl-2004} posited that price impact must decrease over time. This hypothesis has been confirmed by subsequent studies \citep{BouchaudEtAl-2006, Hopman-2007, BouchaudEtAl-2009a, Gatheral-2010, GatheralSchied-2013, AlfonsiEtAl-2016, GarleanuPedersen-2016, TothEtAl-2017, TarantoEtAl-2018, EkrenMuhle-Karbe-2019}. This finding led to the formulation of the propagator model, where prices are expressed as the cumulative impact of all previous trades \citep{BouchaudEtAl-2006,Bouchaud-2009, AlfonsiEtAl-2016, BenzaquenEtAl-2017, BouchaudEtAl-2018, SchneiderLillo-2019}.

A more subtle effect, known as \textit{cross-impact}, occurs when the trading pressure in one asset influences the price of another. This phenomenon was initially studied by \citet{HasbrouckSeppi-2001} and later by \citet{ChordiaEtAl-2001, EvansLyons-2001, HarfordKaul-2005, PasquarielloVega-2007, AndradeEtAl-2008, Tookes-2008, PasquarielloVega-2015, WangGuhr-2017, BenzaquenEtAl-2017, SchneiderLillo-2019, TomasEtAl-2022a, TomasEtAl-2022, BrigoEtAl-2022}. The simplest cross-impact models assume a linear relationship between signed trading volumes and price variations \citep{HasbrouckSeppi-2001, HarfordKaul-2005, PasquarielloVega-2007, PasquarielloVega-2015, TomasEtAl-2022a, TomasEtAl-2022, LeCozEtAl-2024a}. In particular, \citet{LeCozEtAl-2024a} show that the interest rate curve exhibits significant cross-impact features. Bonds of different tenors are highly correlated and display a wide range of liquidity levels, which are the two characteristics required to accurately predict price changes using trading flows. 

In addition, several authors \citep{PlerouEtAl-2000, Cont-2001, BouchaudEtAl-2004, BouchaudEtAl-2018, Elomari-KessabEtAl-2024} have shown that price variations exhibit autocorrelation patterns over short timescales. These findings challenge the traditional viewpoint of market efficiency, which posits that price changes are memoryless.

Here we show that short-time-scale autocorrelation and cross-impact are compatible with the microfounded field theory of the FRC developed in \citet{LeCozBouchaud-2024a}.

\subsection{Definitions and notations}

In this section, we define the forward interest rate and its signed order flow. Table~\ref{tab:notations} in appendix~\ref{Notations} provides a complete list of  notations used in this article.

\subsubsection{Forward interest rate}
Let~$P(t,T)$ denote the price at time~$t$ of a zero-coupon bond maturing at~$T$. Such a bond pays one unit of currency at maturity~$T$ without any intermediate coupons. Consider time~$t$ and a future time~$T$, where $t<T$. The instantaneous forward rate~$f(t,T)$ is defined by
\begin{align}
    f(t,T) = -\frac{\partial \log{P(t,T)}}{\partial T},
\end{align}
The collection of these rates for various~$T$ forms the forward interest rate curve.

In subsequent sections, we define the instantaneous forward rate~$f(t,\theta)$ in terms of the time-to-maturity or \emph{tenor}~$\theta = T-t$. This dimension~$\theta$ is often referred to as the \emph{space} dimension, as opposed to the time dimension~$t$.

\subsubsection{Futures contracts}
The instantaneous forward rate~$f(t,\theta)$ is interpreted as the mid-price at time~$t$ of a 3-month SOFR Futures contract maturing at~$t+\theta$. In practice, Futures contracts are available only for a discrete list of $n$ tenors. In the following sections, we denote any process $x(t)$ defined in the discrete space of the existing tenors as a vector $(x_{\theta}(t))$. Therefore, we depart from the usual notation of the forward rate $f(t,\theta)$ to denote by $f_{\theta}(t)$ the closing forward rate of tenor~$\theta$ in the time window~$[t-\Delta t, t]$ with a length of \(\Delta t = 1\) day. We then denote by $f(t) = (f_{1}(t), \cdots, f_{n}(t) )$ the column vector of the forward rates at closing, with the tenor~$\theta$ in units of $3$-months.

We define $\Delta q_\theta(t)$ as the net market order flow traded during the time window~$[t-\Delta t, t ]$ for the Future contract maturing in~$t+\theta$. This is calculated by taking the sum of the volumes of all trades during that time period, with buy trades counted as positive and sell trades counted as negative. Thus, $\Delta q(t) = (\Delta q_{1}(t), \cdots, \Delta q_{n}(t) )$ is the column vector of net traded order flows.

\subsubsection{Other notations}
The set of real-valued square matrices of dimension~$n$ is denoted by $\mathbf{M}_n(\mathbb{R})$. Given~$A$ a positive symmetric matrix, we write $A^{1/2}$ for a matrix such that $A^{1/2}(A^{1/2})^\top = A$, and $\sqrt{A}$ for the matrix square root: the unique positive semi-definite symmetric matrix such that $(\sqrt{A})^2 = A$. We also write $\diag(A)$ for the vector in~$\mathbb{R}^n$ formed by the diagonal elements of $A$. Given a vector~$v$ in~$\mathbb{R}^n$, we denote the components of $v$ by $(v_1, \cdots, v_n)$, and the diagonal matrix whose components are the components of $v$ by $\diag(v)$. We also define $I_{k}$ with $k \in \llbracket -n,n\rrbracket$ as a matrix with ones only on the $k$-th diagonal above the main diagonal, i.e.,
\begin{equation}
(I_{k})_{ij} =
    \begin{cases}
     1 \quad \text{if} \quad j-i = k, \\
     0 \quad \text{otherwise}.
    \end{cases}
\end{equation}
Note that $I_0$ is the identity matrix simply denoted $I$.

\section{A field theory of the FRC} \label{A field theory of the FRC}
In this section, we summarize several results related to the correlated noise field developed in \citet{LeCozBouchaud-2024a}. We also introduce some additional properties of such a noise field.

Let $\eta(t)$ denote a vector of independent Gaussian (Langevin) noises such that:
\begin{align}
    \E{\eta_{\theta}(t) \eta_{\theta'}(t')}= \delta(t-t') \delta_{\theta\theta'},
\end{align}
where $\delta_{\theta\theta'}$ is the Kronecker delta, $\delta(.)$ is the Dirac delta, and $\E{.}$ denotes an unconditional expectancy. Note that the vector of stochastic processes $B(t)$, defined by ${\rm d}B(t) = \eta(t) {\rm d}t$, represents a multidimensional Brownian motion.

The vector of the driftless discrete noise field~$A(t)$ is defined for $\theta \in \llbracket 1,n\rrbracket$ as the solution to a differential equation which operates on a temporal scale~$\tau \ll 1$~day:
\begin{equation}
    \label{eq:discret_master_general}
    \left\lbrace
    \begin{aligned}
   & \frac{{\rm d}A}{{\rm d}t}(t)  = \frac{1}{\tau} \left[ -\mathcal{M} A(t)+\eta(t)\right], \\
    &A_{1}(t) - A_{-1}(t)= 0,
    \end{aligned}
    \right.
\end{equation}
where $\mathcal{M}$ is a matrix of $\mathbf{M}_n(\mathbb{R})$ defined by
\begin{align}
    \mathcal{M}_{\theta\theta'}=1 - \frac{1}{2\psi\mu^2} \left(1+\frac{\theta}{\psi}\right)\left(I_{1} - I_{-1} \right)_{\theta\theta'} \nonumber \\ 
    - \frac{1}{\mu^2}\left(1+\frac{\theta}{\psi}\right)^2\left(I_{1} -2 I + I_{-1} \right)_{\theta\theta'},
\end{align}
with $\psi$ the psychological time parameter and $\mu$ the line tension parameter \citep{LeCozBouchaud-2024a}. Note that the boundary condition in Eq.~\eqref{eq:discret_master_general} exhibits a term $A_{-1}$ generated by the use of an Euler scheme centered in $0$ to ensure the validity of the method of images used in \citet{LeCozBouchaud-2024a}.

 Here, we use a white noise $\eta(t){\rm d}t$ of variance ${\rm d}t$ instead of $2D{\rm d}t$ as in~\citet{LeCozBouchaud-2024a};  this choice has no impact on the results. Moreover, we consider a finite number $n$ of diffusion factors, one for each tenor of the FRC, although the model could be written with an infinite-dimensional white noise $\eta$. In any case, only the first $n$ component of the vector $Y$ (see section~\ref{Noises decomposition}) would be non-zero. 

Although Eq.~\eqref{eq:discret_master_general} cannot be solved in closed form for arbitrary values of $\psi$, it simplifies in the two limits $\psi \to \infty$ and $\psi \to 0$. The general solution to Eq.~\eqref{eq:discret_master_general} is expressed as
\begin{align}
    \label{eq:corr_noise_def}
    A(t) =  \frac{1}{\tau} \int_{-\infty}^t {\rm d}t' G(t-t') \eta(t'),
\end{align}
where the matrix $G(t-t')$ is the \textit{propagator} of the noise $\eta(t')$. When $\psi \gg 1$, for $(\theta,\theta') \in \llbracket 1,n\rrbracket^2$, $G_{\theta\theta'}$ is given by \citep{LeCozBouchaud-2024a}:
\begin{align}
    &G_{\theta\theta'}(t) := \nonumber \\
    & \frac{1}{2\pi}\int_{-\pi}^{\pi} {\rm d}\xi \left( e^{i\xi(\theta-\theta')} + e^{i\xi(\theta+\theta')} \right) e^{-\frac{L_{d}(\xi)}{\tau}t},
\end{align}
where $L_{d}(\xi) = 1 + 2\frac{(1-\cos{\xi})}{\mu^2}$. When $\psi \ll 1$, $G$ becomes \citep{LeCozBouchaud-2024a}:
\begin{align}
    G(t) := e^{-\frac{t}{\tau}\mathcal{M}} \mathcal{J},
\end{align}
where $\mathcal{J}$ denotes a diagonal matrix whose first entry is $2$ and all other entries are $1$. In this limit, the matrix $\mathcal{M}$ can be written as a function of a single parameter $\kappa = \mu\psi$:
\begin{align}
    \label{eq:curvy_M}
    \mathcal{M}_{\theta\theta'}=I - \frac{\theta}{\kappa^2}\left( I_{1} - I_{-1} \right) - \frac{\theta^2}{\kappa^2}\left(I_{1} -2 I + I_{-1}\right).
\end{align}

A critical characteristic of the noise field~$A(t,\theta)$ is its auto-covariance across time and space. For~$\tau$ near $0$, the auto-covariance of its integral over a time interval~$\Delta t$, defined by~$\Delta A := \int_{t-\Delta t}^t A_{\theta}(u) \mathrm{d}u$ is given by \citep{LeCozBouchaud-2024a}:
 \begin{equation}
    \E{ \Delta A(t) \Delta A^\top(t')} =
    \left\lbrace
    \begin{aligned}
    &0, \text{ if } |t-t'|> \Delta t, \\
    &  \Delta t \; C, \text{ if } t=t',
    \end{aligned}
    \right.
\end{equation}
where the matrix $C$ is the \textit{correlator} of $\Delta A$. The latter is defined by \citep{LeCozBouchaud-2024a}:
\begin{align}
\label{eq:correlator_definition}
C := \begin{cases}
    \mathcal{D}_2 &\text{ if } \psi \gg 1, \\
    \mathcal{M}^{-1}\mathcal{J}^2(\mathcal{M}^{-1})^\top &\text{ if } \psi \ll 1,
\end{cases}
\end{align}
where the matrix $\mathcal{D}_k$ is given by
\begin{align}
    \label{eq:mathcal_D_definition}
    (\mathcal{D}_k)_{\theta\theta'} := \frac{1}{\pi}\int_{0}^{\pi}{\rm d}\xi \frac{2\cos{\xi \theta}\cos{\xi \theta'}}{L_{d}(\xi)^k}.
\end{align}

The second important property of the cumulative sum of~$A$ is its response to the generating white noise. We define the integrated $\eta$ by $\Delta \eta_{\theta}(t) := \int_{t-\Delta t}^t \eta_{\theta}(u) {\rm d}u$.  For~$\tau$ near $0$, the covariance between $\Delta A(t)$ and $\Delta \eta(t)$ reads
\begin{align}
    & \E{\Delta A(t) \Delta \eta^\top(t)} =  \Delta t \; R,
\end{align}
where the matrix $R$ is the response of $\Delta A$ to $\Delta \eta$ given by:
\begin{align}
R := \begin{cases}
    \mathcal{D}_1 &\text{ if } \psi \gg 1, \\
    \mathcal{M}^{-1}\mathcal{J} &\text{ if } \psi \ll 1.
\end{cases}
\end{align}
Hence, the correlation matrix $\rho(\Delta A(t),\Delta \eta(t))$ between $\Delta A(t)$ and $\Delta \eta(t)$ reads
\begin{align}
    \label{eq:corr_DeltaA_Deltaeta}
    \rho(\Delta A(t),\Delta \eta(t)) = \diag(\sigma_A)^{-1}R, 
\end{align}
where $\sigma_A$ is the volatility vector of $\Delta A$ defined by
\begin{align}
    (\sigma_A)_\theta = \sqrt{\diag(C)_\theta}.
\end{align}
The proofs of these properties are provided in appendices~\ref{Response to the white noise for psi above 1} and~\ref{Response to the white noise for psi below 1}.

The noise field $A$ is now employed to model forward rates. The diffusion equation for the variations of the forward rate, denoted as~${\rm d}f_{\theta}(t)$, is given by \citep{LeCozBouchaud-2024a}:
\begin{align}
    \label{eq:forward_rate_diffusion}
    \frac{{\rm d} f}{{\rm d} t} (t) = \diag(\sigma) \diag(\sigma_A)^{-1} A(t),
\end{align}
where the component $\sigma_{\theta}$ of the vector $\sigma$ is the volatility of the noise term driving the forward rate $f_\theta$. Hence, the equal-time Pearson correlation coefficient among coarse-grained forward rate variations $\Delta f := \int_{t-\Delta t}^t {\rm d} f(u)$ is given by \citep{LeCozBouchaud-2024a}:
\begin{align}
    \label{eq:micro-founded_final}
    \rho(\Delta f,\Delta f) = \diag(\sigma_A)^{-1} C \diag(\sigma_A)^{-1}.
\end{align}

\section{Towards a cross-impact model}
Although the model of \citet{LeCozBouchaud-2024a} provides an accurate account of the correlation structure of the FRC, it does not clarify the nature of the exogenous noise $\eta$ driving the dynamics of the curve. In this section we want to provide a microstructural foundation for the $\eta$ noise by linking it to the surprise in the order flow, thus promoting the model (that only describes price variations) to a microstructural model accounting for the joint dynamics of prices and volumes.

\subsection{Order flow decomposition}
\begin{figure}
\centering
 \includegraphics[width=\linewidth]{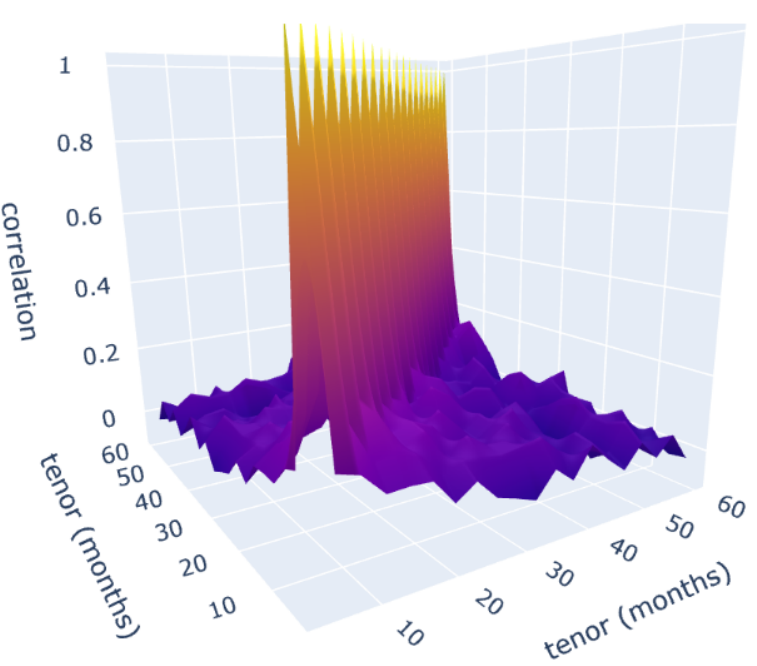}
 \caption{Spatial Pearson correlations of the signed daily order flows of SOFR Futures from $2016$ to $2023$.} 
\label{fig:3dplot_correlation_volumes_screen_shot}
\end{figure}

Trading flows exhibit significantly lower spatial correlation compared to prices, as shown in Fig.~\ref{fig:3dplot_correlation_volumes_screen_shot}. 
However, they display long range temporal autocorrelation (see \citet{BouchaudEtAl-2018} and the literature mentioned in section~\ref{Market micro-structure}). Therefore, we provide a natural physical interpretation of the white noise $\eta$ by assuming that this noise corresponds to the surprise (i.e.~the martingale component) in the signed order flow. However, only a fraction of the volatility of price increments is expected to be explained by trades (see section~\ref{Asymmetric responses between flows and prices}), so the white noise column vector $\eta(t)$ is decomposed into an idiosyncratic component $\eta^{\perp}$, and a component related to order flow $\eta^q$:
\begin{align}
    \label{eq:eta_flow}
    & \eta(t) = \diag(Y) \eta^{q}(t) + \diag(Y^{\perp}) \eta^{\perp}(t),
\end{align}
where $\eta^{\perp}$ is a normalized white noise independent from $\eta^q$ and $Y$ is the vector of the parameters $Y_\theta \in [0,1]$ governing, for each tenor $\theta$, the share of forward rates variance explained by the order flow imbalance.  The  components of the vector $Y^{\perp}$ are $Y_\theta^{\perp}  = \sqrt{1-Y_\theta^2}$.
Formally, the surprise $\eta^q(t)$ is defined as
\begin{align}
    \label{eq:eta_q_definition}
    \eta^{q}(t) &:= O \int_{-\infty}^t{\rm d}t' J(t-t') \frac{{\rm d}q}{{\rm d}t}(t'),
\end{align}
where $\frac{{\rm d}q}{{\rm d}t}(t')$ is the infinitesimal order flow imbalance, and $J(t-t')$ is a matrix-valued function that ensures the diffusivity of the process $\eta^{q}(t)$, i.e., $\E{\eta^{q}_{\theta}(t)\eta^{q}_{\theta'}(t')} = \delta_{\theta\theta'} \delta(t-t')$. Such an operator is defined up to an arbitrary rotation matrix $O$ which leaves the price process invariant.

In appendix~\ref{app:Order flow decomposition}, we justify the existence of the kernel~$J$ in Eq.~\eqref{eq:eta_q_definition}, assuming the lagged variance-covariance matrix of the infinitesimal order flows~$\Omega(t,t'):=\E{\frac{{\rm d}q}{{\rm d}t}(t) \frac{{\rm d}q}{{\rm d}t}^{\top}(t')} $ is stationary: 
$
    \Omega(t,t') = \Omega(t-t').
$
We further assume that the order flow has a factorized structure\footnote{Even though this assumption is not strictly required in our construction, we prefer to stick to this simpler case, which is an acceptable first order approximation of the empirical order flow structure, see~\citep{BenzaquenEtAl-2016b}.}
\begin{align}
   \Omega(t-t') = \diag(\phi(t-t')) \Omega,
\end{align}
where the function $\ell \mapsto \phi(\ell)$ is valued in vector space, and $\Omega$ is the equal-time variance-covariance matrix of the infinitesimal order flows. Then, it is quite simple to obtain an explicit expression for $J(t-t')$ (see appendix~\ref{app:Order flow decomposition}):
\begin{align}
    J(t-t') = \Omega^{-1/2} \diag{( \Phi(t-t'))},
\end{align}
where $\Phi(\ell)$ is an operator, valued in vector space, denoting the element-wise convolutional inverse of $\phi(\ell)$.

Note that this construction leaves the rotation matrix $O$ undefined. In section~\ref{Cross-impact matrix}, we propose a method to determine this matrix to satisfy the consistency requirements of a cross-impact model.

Finally, one can define $\tilde{q}(t)$, the martingale component of $q(t)$, through:
\begin{align}
    \frac{{\rm d}\tilde{q}}{{\rm d}t}(t) := \int_{-\infty}^t{\rm d}t'\diag{( \Phi(t-t'))} \frac{{\rm d}q}{{\rm d}t}(t'),
\end{align}
such that the surprise $\eta^q$ is given by
\begin{align}
    \label{eq:white_noise_building}
    \eta^{q}(t) &= O \Omega^{-1/2} \frac{{\rm d}\tilde{q}}{{\rm d}t}(t).
\end{align}

\subsection{Noise field decomposition} \label{Noises decomposition}

The decomposition of the white noise~$\eta$ enables us to write the noise field~$A$ as the sum of two independent components:
\begin{align}
A(t) = A^q(t) + A^{\perp}(t),
\end{align}
where the correlated noise $A^q$ is the solution of
\begin{equation}
    \label{eq:master_flow}
    \left\lbrace
    \begin{aligned}
    &\frac{{\rm d} A^q}{{\rm d} t}(t) = \frac{1}{\tau} \left[ -\mathcal{M}A^q(t) +  \diag(Y) \eta^q(t) \right], \\
    &A^q_{1}(t) - A^q_{-1}(t)= 0.
    \end{aligned}
    \right.
\end{equation}
The correlated noise $A^{\perp}$ solves a similar equation with the generating white noise $\diag(Y^{\perp}) \eta^{\perp}(t)$.

\subsection{Large-bin approximation}
Even though Eq.~\eqref{eq:corr_noise_def} shows that $\Delta A(t)$ depends upon the whole history of $\eta(t')$ for $t'\le t$, we are interested in approximating $\Delta A_\theta(t)$ as a function of coarse-grained variables $\Delta \eta$ defined over intervals of finite width $\Delta t$, as in practice we will have empirical access to order flows sampled on a discrete time grid. The proofs of the results presented in this section are detailed in appendix~\ref{appendix large-bin approximation}.

We decompose the white noise column vector $\eta(t')$ into the sum of its observed empirical averages over the time intervals $[t - \Delta t,t]$ (i.e. its moving average) and its fluctuations around this mean. Formally, we write
\begin{align}
    \label{eq:eta_decomposition}
    \eta(t') = \bar{\eta}_{\Delta t}(t) + \eta(t') - \bar{\eta}_{\Delta t}(t),
\end{align}
where  $\bar{\eta}_{\Delta t}(t)$ is the empirical mean of $\eta(t)$ over the time window $[t - \Delta t,t]$ i.e.,
\begin{align}
    \bar{\eta}_{\Delta t}(t) &:= \frac{1}{\Delta t}\int_{t-\Delta t}^t {\rm d}t' \eta(t').  
\end{align}
If we further consider that $\tau \ll \Delta t$ one can express $\Delta A_\theta(t)$ as a function of $\bar{\eta}_{\Delta t}(t)$:
\begin{align}
    \label{eq:delta_A_lin_main}
    \Delta A(t) = R \; \Delta t \; \bar{\eta}_{\Delta t}(t) + \int_{t-\Delta t}^{t} {\rm d}t' \epsilon^{\tau}(t'),
\end{align}
where $\epsilon^{\tau}(t) = \frac{1}{\tau} \int_{t-\Delta t}^t {\rm d}t' G(t-t') \left( \eta(t') - \bar{\eta}_{\Delta t}(t) \right)$ is a noise independent of $\bar{\eta}_{\Delta t}(t)$ (see appendix~\ref{appendix large-bin approximation}). One can substitute $\eta$ with $\eta^q$ or $\eta^{\top}$ and $A$ with $A^q$ or $A^{\top}$ in Eq.~\eqref{eq:delta_A_lin_main}. It yields a relationship between forward rate daily increments and the martingale component of the daily order flow~$\Delta \tilde{q}(t):= \int_{t-\Delta t}^t {\rm d}t'  \tilde{q}(t')$:
\begin{align}
     \widehat{\Delta f}(t) & = \diag(\sigma) \diag(\sigma_A)^{-1} R \diag(Y) O \Omega^{-1/2}\Delta \tilde{q}(t),
\end{align}
where $\widehat{\Delta f}$ denotes the conditional expectancy of the forward rates increments~$\Delta f$ with respect to these flows: 
\begin{align}
    \widehat{\Delta f}(t) := \mathbb{E}\left[ \Delta f(t) \right| \left. \Delta \tilde{q}(t) \right].
\end{align}

\begin{figure}
\centering
 \includegraphics[width=0.85\linewidth]{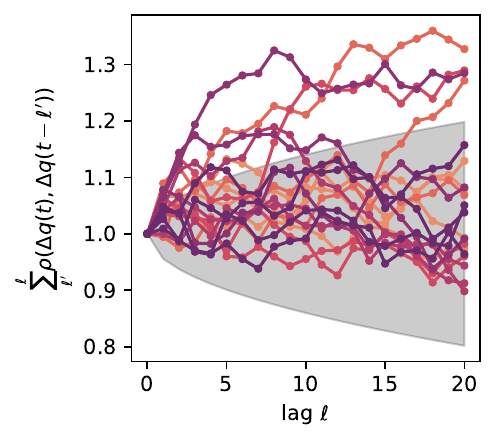}
 \caption{Accumulated temporal autocorrelation of daily trading flows over $\ell$ days i.e., $\sum_{\ell'}^{\ell} \rho(\Delta q(t),\Delta q(t-\ell'))$. Each color corresponds to the tenor of a SOFR Future contract ranging from $3$ to $60$ months over the period $2016-2023$. Only $4$ maturities ($15$, $21$, $27$ and $45$ months) out-of $20$ are outside the confidence interval after 20 days.} 
 \label{fig:Delta_q_temporal_correlation}
\end{figure}

{Similarly to the approach of \citet{LeCozEtAl-2024a}, we neglect the autocorrelation of the order flows}, such that $\Delta \tilde{q}(t) \approx \Delta q(t)$. This approximation is adequate on the daily time scale for $80\%$ of the maturities considered in our sample (see  Fig.~\ref{fig:Delta_q_temporal_correlation}).
Hence, one can write the conditional expectancy of the forward rates daily increments with respect to the order flow as
\begin{align}
    \label{eq:calibration_Y}
     \widehat{\Delta f}(t) = \diag(\sigma) \diag(\sigma_A)^{-1} R \diag(Y) O \Omega^{-1/2}\Delta q(t).
\end{align}

\subsection{Cross-impact matrix} \label{Cross-impact matrix}
Equation~\eqref{eq:calibration_Y} can be used to define a cross-impact model. Let $\Lambda \in \mathbf{M}_n(\mathbb{R})$ be the matrix such that the equal-bin linear relationship between forward rates increments and order flows reads
\begin{align}
    \label{eq:definition_lambda}
    \Delta f(t) = \Lambda \Delta q(t) + \mathcal{E}(t),
\end{align}
where $\mathcal{E}$ is a temporally uncorrelated noise independent from $\Delta q$. Identifying $\Lambda$ in Eq.~\eqref{eq:calibration_Y} yields
\begin{align}
    \label{eq:lambda_def}
    \Lambda = \diag(\sigma) \diag(\sigma_A)^{-1} R \diag(Y) O \Omega^{-1/2}.
\end{align}
This formula can also be derived by computing $\Lambda$ as the linear response of the forward rates to the equal-time order-flow:
\begin{align}
    \Lambda := \E{\Delta f(t) \Delta q(t)^\top}\E{\Delta q(t) {\Delta q(t)}^\top}^{-1}.
\end{align}
The proof of this alternative approach is provided in appendix~\ref{Response to order flows for psi above 1}. In addition, appendix~\ref{Correlation between forward rates and flows} shows that the correlation between the forward rate and the order flow is well defined.

The cross-impact model in Eq.~\eqref{eq:lambda_def} is fully determined up to an arbitrary rotation matrix $O$. This free parameter can be used to ensure that our model has the required (i) rotational invariance, (ii) non-arbitrage, (iii) fragmentation invariance, and (iv) stability properties \citep{TomasEtAl-2022}. In fact, it was shown that the cross-impact matrix $\Lambda$ that satisfies these properties must be symmetric positive definite \citep{TomasEtAl-2022}. The rotation $O_{\text{sym}}$ ensuring that $\Lambda$ fulfills these properties is given by \citep{delMolinoEtAl-2020}:
\begin{align}
    \label{eq:O_symetrical}
    O_{\text{sym}}(M,\Omega^{1/2}) & := M^{-1} (\Omega^{-1/2})^\top \nonumber \\
    & \hspace{1cm} \sqrt{(\Omega^{1/2})^\top M M^{\top}\Omega^{1/2}},
\end{align}
where $M = \diag(\sigma) \diag(\sigma_A)^{-1} R \diag(Y)$.

As an alternative model that does not meet these constraints, one can also simply choose $O=I$ the identity matrix. We will refer to the cross-impact model using Eq.~\eqref{eq:lambda_def} as BBDLW for Baaquie-Bouchaud Discrete Logarithm Whitening when $O=I$ and BBDLS for Baaquie-Bouchaud Discrete Logarithm Symmetric when $O=O_{\text{sym}}$. 

\section{Calibration}
The best empirical fits of the BB model are obtained in the limit $\psi \to 0$ (see \citet{LeCozBouchaud-2024a}), which is therefore chosen for calibration.

Our data set comprises historical daily price variations and net market order flows of SOFR Futures contracts from July $2015$ to $2023$. We observe $n=20$ different tenors ranging from $3$ to $60$ months. $3$-month SOFR Futures contracts were not available before March 2022; thus, Eurodollar contracts were used before that time, with an appropriate three-month shift accounting for the forward-looking nature of the Eurodollar Futures as opposed to the backward-accrued SOFR.

\subsection{Methodology} \label{Methodology}

In line with the approach of \citet{LeCozBouchaud-2024a}, we fit the parameter $\kappa$ in formula~\eqref{eq:micro-founded_final} to the observed forward rate correlation matrix within our dataset segmented into $3$ periods: $2015-2017$, $2018-2020$ and $2021-2023$. In addition, we fit the vector $Y$ by minimizing the square differences between the daily increments of modeled forward rates, $\widehat{\Delta f}$, and the empirical ones, $\Delta f$, using Eq.~\eqref{eq:calibration_Y}.

To overcome the conditional heteroskedasticity of forward rate variations, we use a daily estimator of their volatility. Let~$\{t_1, \cdots, t_N\}$ denote the $N$ business days of a period of $3$ years. For each day~$t_k$, the estimators of the forward rates increments and order flow's volatility are defined by
\begin{align}
\widehat{\sigma}^2(t_k) := (\langle {\Delta f_{1}(t)}^2 \rangle(t_k), \cdots, \langle{\Delta f_{n}(t)}^2\rangle(t_k)), \nonumber \\
\widehat{\omega}^2(t_k) := (\langle {\Delta q_{1}(t)}^2 \rangle(t_k), \cdots, \langle{\Delta q_{n}(t)}^2\rangle(t_k)),
\end{align}
where the operator $\langle.\rangle(t_k)$ denotes the moving-average computed using the last $20$ daily data points before the day~$t_k$. We assume the order flow correlation matrix~$\diag({\omega}(t))^{-1} \Omega(t) \diag({\omega}(t))^{-1}$ is stationary. Let~$\widehat{\rho}(\Delta q,\Delta q)$ denote its canonical empirical estimator using $3$ years of data. On day~$t_k$, the estimated variance-covariance matrix~$\widehat{\Omega}(t_k)$ is computed as
\begin{align}
    \widehat{\Omega}(t_k) := \diag(\widehat{\omega}(t_k)) \widehat{\rho}(\Delta q,\Delta q) \diag(\widehat{\omega}(t_k)).
\end{align}
We define similarly the estimated variance-covariance  of the forward rate variations and the estimated response matrices as
\begin{align}
    \widehat{\Sigma}(t_k) := \diag(\widehat{\sigma}(t_k)) \widehat{\rho}(\Delta f,\Delta f) \diag(\widehat{\sigma}(t_k)), \nonumber \\
    \widehat{R}(t_k) :=  \diag(\widehat{\sigma}(t_k)) \widehat{\rho}(\Delta f,\Delta q) \diag(\widehat{\omega}(t_k)).
\end{align}
 The predicted forward rate change on day~$t_k$ is defined as
\begin{align}
    \widehat{\Delta f}(t_k) &= \widehat{\Lambda}^{\text{model}}(t_k)\Delta q(t_k),
\end{align}
where $\widehat{\Lambda}^{\text{model}}(t_k)$ is the cross-impact matrix estimated on day~$t_k$ in the tested model.

In the case of our noise field approach, the cross-impact matrix is given by
\begin{align}
\label{eq:BB_cross_impact_matrix}
    \widehat{\Lambda}^{\text{BB}}(t_k) = \diag(\widehat{\sigma}(t_k)) \diag(\sigma_A)^{-1} R \diag(Y) O \widehat{\Omega}(t_k)^{-1/2},
\end{align}
where $O=O_{\text{sym}}$ or $O=I$ depending on the tested model.

In order to compare the results of our model with other cross-impact models, we define three other cross-impact matrices (studied, for example, in \citet{LeCozEtAl-2024a}). Let $y$ denote a scalar called the Y-ratio. We consider:
\begin{itemize}
\item the diagonal model, defined by
\begin{align}
    \widehat{\Lambda}^{\text{diag}}(t_k) := y \diag(\widehat{R}(t_k))\diag(\widehat{\Omega}(t_k)^{-1}),
\end{align}
which is the limit case where the cross-sectional impact is set to zero;

\item the Maximum Likelihood model (ML model in the following sections), defined by
\begin{align}
    \widehat{\Lambda}^{\text{ML}}(t_k) := y \widehat{R}(t_k) \widehat{\Omega}(t_k)^{-1};
\end{align}

\item and the so-called Kyle model, defined by
\begin{align}
    &\widehat{\Lambda}^{\text{Kyle}}(t_k) := \nonumber \\
    & \hspace{1cm} y \widehat{\Sigma}(t_k)^{1/2} O_{\text{sym}}\Big(\widehat{\Sigma}(t_k)^{1/2},\widehat{\Omega}(t_k)^{1/2}\Big) \widehat{\Omega}(t_k)^{-1/2}.
\end{align}
    
\end{itemize}

The ML model does not impose any constraints on the cross-impact model, so it generates the best possible in-sample fit. The Kyle model ensures (i) rotational invariance, (ii) non-arbitrage, (iii) fragmentation invariance, and (iv) stability properties \citep{TomasEtAl-2022, delMolinoEtAl-2020}. However, none of these models prescribes the form of the price variance-covariance matrix. Such a matrix is fully determined within the BBDL model thanks to a single parameter $\kappa$ in the case $\psi \ll 1$ (see section~\ref{A field theory of the FRC}).

\subsection{Goodness-of-fit}
To assess the model goodness-of-fit, we compare the predicted price changes~$\widehat{\Delta f}(t)$ with the realized price changes~$\Delta f(t)$. For this evaluation, we employ a generalized R-squared, parameterized by a symmetric, positive matrix~$W$. The $W$-weighted generalized $\mathcal{R}^2(W)$ is given by
\begin{align}
    &\mathcal{R}^2(W) := \nonumber \\
    &1 - \frac{\Sigma_{k=1}^N\left(\Delta f(t_k)-\widehat{\Delta f}(t_k)\right)^\top W(t_k) \left(\Delta f(t_k)-\widehat{\Delta f}(t_k)\right)}{\Sigma_{k=1}^N\Delta f(t_k)^\top W(t_k) \Delta f(t_k)}.
\end{align}
The closer this score is to one, the better the fit to actual prices. To highlight different sources of error, different choices of $W$ can be considered:
\begin{itemize}
\item $W_{\sigma}(t)  := {\diag(\widehat{\sigma}^2(t))}^{-1}$, to account for errors relative to the typical deviation of the assets considered. This type of error is relevant for strategies that predict idiosyncratic moves of the constituents of the basket, rather than strategies that wager on correlated market moves.
\item $W_{\sigma_\theta}(t) := {\diag(( 0,\dots,0,\widehat{\sigma}^2_{\theta}(t)},0,\dots,0))^{-1}$, to account for the errors of a single asset~$\theta$.
\end{itemize}
The weights $W_{\sigma}$ are used in section~\ref{Results} to compare the overall performance of cross-impact models, while the weights $W_{\sigma_\theta}(t)$ are used in section~\ref{Asymmetric responses between flows and prices} to measure their properties in a pairwise setting. 

\subsection{Results} \label{Results}
The results of the calibration of the BBDL model on empirical correlations of the forward rate are presented in table~\ref{tab:calibration_on_prices}. This confirms the high accuracy and good parameter stability of the BBDL model.
 \begingroup
\begin{table}[t]
    \begin{ruledtabular}
    \begin{tabular}{ccccc}
    Period & $\kappa$  & $\mathcal{R}^2$\\
    \hline
    2015--2017 & 0.84  & 99.9\%\\
    2018--2020 & 0.82  & 99.4\%\\
    2021--2023 & 1.3  & 97.1\%\\
    \end{tabular}
    \end{ruledtabular}
    \caption{Calibrated line tension parameter $\kappa$ in the BBDL model for each $3$-year period in our sample. Here, $\mathcal{R}^2$  denotes the share of the explained variance of the empirical correlations among forward rates of time-to-maturity ranging from $3$ to $60$ months.}
    \label{tab:calibration_on_prices}
\end{table}
\endgroup

\begin{figure}
\centering
 \includegraphics[width=0.85\linewidth]{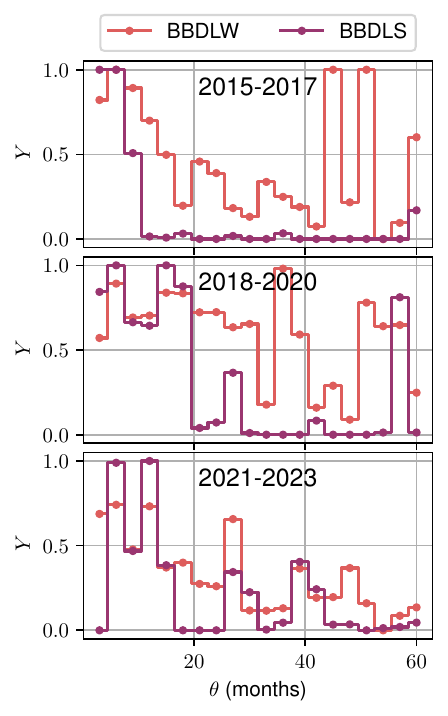}
 \caption{Optimal parameters $Y_\theta$ governing the share of forward rates variances explained by order flows for each maturity $\theta$.} 
 \label{fig:Y_star_plot}
\end{figure}

 Using the calibrated line tension parameter~$\kappa$ reported in table~\ref{tab:calibration_on_prices}, we fit the share of explained volatility $Y$ to the time series of rates and order flows of SOFR Futures. The calibrated share $Y$ of the explained forward rate volatility is reported for each period in Fig.~\ref{fig:Y_star_plot}\footnote{More precisely, this calibration is performed by assuming $Y \in [0,1]$ and calibrating the \textit{prices} of SOFR Futures $p(t, \theta) = 100 - f(t, \theta)$ to the signed order flow. If we relax this constraint, the search for values $Y \in [-1,1]$ also yields (almost) systematically positive coefficients because of the well-documented positive correlation between order flow and prices \citep{LeCozEtAl-2024a}. Indeed, across all periods and maturities ($60$ calibrated coefficients), we observe only $2$ parameters $Y_{\theta}$ with slightly negative values (around $-0.1$).}. It shows that the most liquid products (the shortest time-to-maturity~$\theta$) are associated with the highest values of $Y_\theta$. We also observe that building the symmetric cross-impact model BBDLS requires setting the less liquid maturities of $Y$ to zero (see Fig.~\ref{fig:Y_star_plot}) in order to avoid instabilities. One could improve the R-squared by putting more weight on the non-liquid products, but this would compromise the model's no-arbitrage property.

\begin{figure}
\centering
 \includegraphics[width=0.75\linewidth]{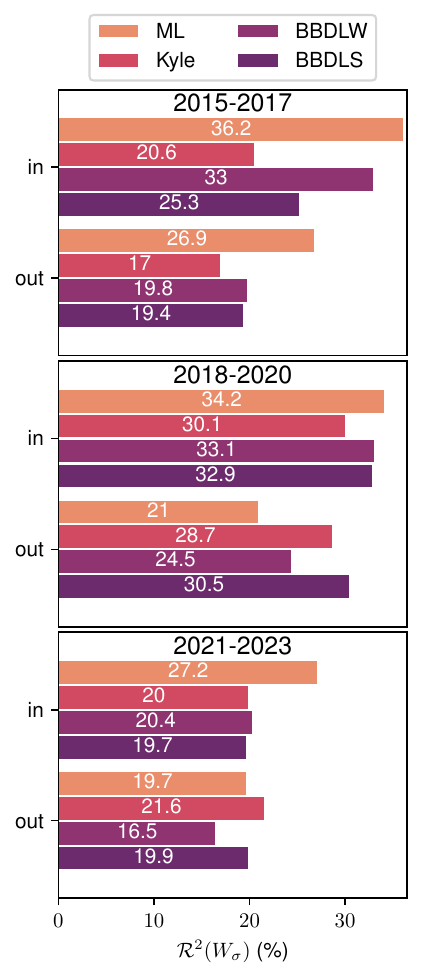}
 \caption{Generalized R-squared $\mathcal{R}^2(W_{\sigma})$ values both in- and out-sample for each period and model. The out--sample values are derived using the parameters calibrated from the preceding period. For the first period $2015-2017$, the out--sample R-squared values are calculated using the parameters calibrated from the $2021-2023$ period.} 
 \label{fig:R2_in_out_all_periods}
\end{figure}

The in-sample R-squared values reported in Fig.~\ref{fig:R2_in_out_all_periods} show that, as expected, the BBDLW model performs worse than the unconstrained ML model, whereas the out-of-sample results show that the two models have similar performances. It is noteworthy because the BBDLW model uses only $n+1$ parameters (excluding the parameters used for the whitening of the correlation matrix), while the ML model requires calibrating $n^2$ parameters. Furthermore, the BBDLS model is actually at least as accurate as the Kyle model in predicting forward rate moves. Once again, this is remarkable because the BBDLS model is more parsimonious than the Kyle model, which uses $\frac{n(n+1)}{2}$ parameters. In fact, the Kyle model, which features a unique Y-ratio for all assets, cannot explore the regime probed by BBDLS.

As expected, compared to the less constrained approaches (ML and BBDLW), the symmetrical models (Kyle and BBDLS) perform poorly in-sample (see Fig.~\ref{fig:R2_in_out_all_periods}). However, the out-of-sample results favor these symmetrical models, demonstrating that the market does not feature arbitrage opportunities large enough to rule out those models.

\begin{figure}
\centering
 \includegraphics[width=0.9\linewidth]{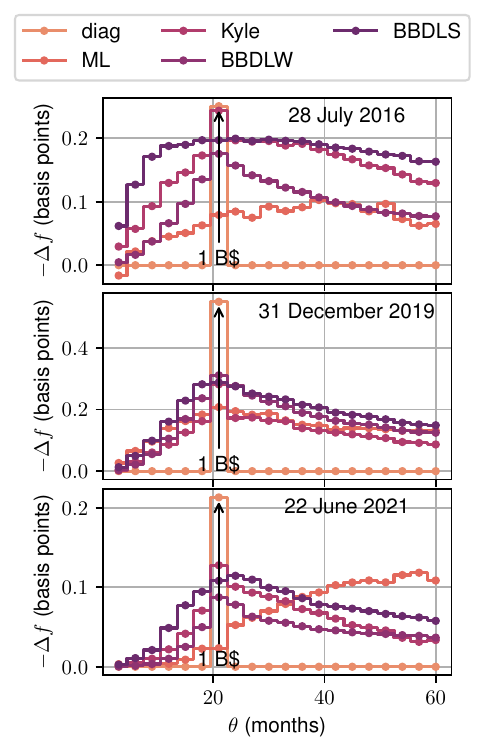}
 \caption{Modeled FRC moves in response to a trade in the SOFR Future of maturity $21$ months with a notional value of one billion dollars over the course of one  randomly chosen day for each of the three calibration periods.} 
 \label{fig:marketing_plot_general}
\end{figure}

Using this set of calibrated parameters, one can draw the response of the forward rate curve to a trade with a notional value of one billion dollars in a single maturity over the course of one day. As our cross-impact models are re-scaled everyday by the daily volatility of prices and order flows (see section~\ref{Methodology}), Fig.~\ref{fig:marketing_plot_general} presents one column of the matrix $\Lambda$ on a randomly chosen day within each three-year period. It shows that all models except the ML model predict a high price impact for the tenor being bought (in this case, $21$ months).

\section{Non-martingality at small time-scales}

Several authors have shown that price variations exhibit autocorrelation patterns  over short time intervals (see the literature cited in section~\ref{Market micro-structure}). Our model is compatible with these results because, for time-scales ${\rm d} t \leq \tau$ the forward rate process is not yet a martingale.

To illustrate this phenomenon, we assume that a volume $V$ of the SOFR Future of time-to-maturity $\theta_0$ is purchased during the time interval ${\rm d} t$. Our goal is to calculate the progressive deformation of the FRC in response to this single trade. Formally, we define ${\rm d}\tilde{q}(0) = V_{\theta_0}$, where $V_{\theta_0}= (0,\dots,0, V, 0, \dots,0)$ is a vector with a single non-zero component $V$ in the position $\theta_0$ and ${\rm d}\tilde{q}(t) = 0$ for $t>0$.  We now discretize time in Eqs~\eqref{eq:discret_master_general}, \eqref{eq:forward_rate_diffusion}, \eqref{eq:white_noise_building}, and~\eqref{eq:master_flow}. This yields an expression for the forward rate variations ${\rm d} f$ at discrete times $k {\rm d} t$ in response to this single transaction:
\begin{align}
    {\rm d} f (k {\rm d} t) = {\rm d} \widehat{f} (k{\rm d} t) + \epsilon'(k\rm{d} t).
\end{align}
Here, $\epsilon'(k\rm{d} t)$ is a noise independent from the forward rate variations  caused by trading activity ${\rm d} \widehat{f} (k{\rm d} t)$, which is defined as ${\rm d} \widehat{f} (k{\rm d} t) :=$
\begin{equation}
\frac{1}{\tau}\diag(\sigma) \diag(\sigma_A)^{-1} \left(I  - \frac{{\rm d} t}{\tau}\mathcal{M}\right)^k \diag(Y) \Omega^{-1/2} V_{\theta_0}.
\end{equation}

\begin{figure}
\centering
 \includegraphics[width=0.85\linewidth]{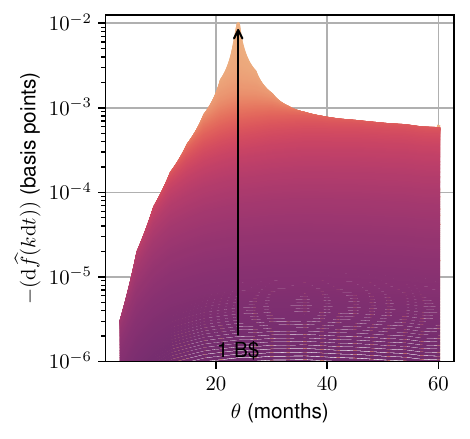}
 \caption{FRC moves in the BBDLW model in response to a trade in the SOFR Future of maturity $24$ months with a notional value of one billion dollars executed at time $t=0$. Each color corresponds to a time step ordered from $t=0.25\% \times \tau$ (orange) to $3\tau$ (purple). The total response $\Delta \widehat{f}$ over the time interval $\Delta t = 3\tau$ is the sum of all the infinitesimal responses ${\rm d} f (k {\rm d} t)$.} 
 \label{fig:impact_across_time_kyle}
\end{figure}

 Fig.~\ref{fig:impact_across_time_kyle} shows the predicted FRC responses to a transaction of volume $V=1$ billion dollars in the $24$-month Future occurring between $t=0$ and $t=0.25\% \times \tau$. Immediately following the trade, the price impact peaks at the traded maturity. The effect of this trade on the other tenors progressively spreads up to $3\tau$, where it becomes negligible. Based on the calibration by \citet{LeCozBouchaud-2024a}, $\tau \approx 30$ minutes. Thus, Fig.~\ref{fig:impact_across_time_kyle} represents the resulting deformation of the FRC between $5$ seconds and $1.5$ hours after the trade.

\section{Influence of liquidity on price-volume correlations} \label{Asymmetric responses between flows and prices}
\begin{figure}
\centering
 \includegraphics[width=\linewidth]{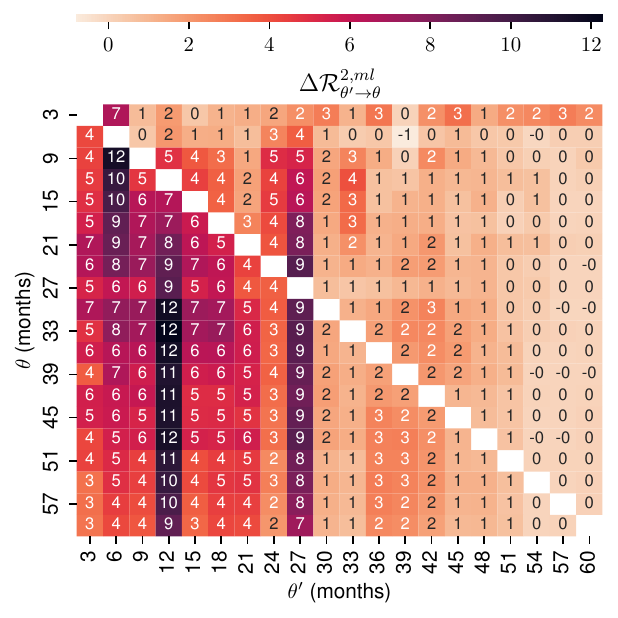}
 \caption{Empirical added accuracy $\Delta \mathcal{R}^{2}_{\theta'\to\theta}$ when regressing daily forward rate increments of tenor $\theta$ on the order flows of tenor $\theta$ and $\theta'$ instead of solely on its own order flow, for the period $2021-2023$. The calibration methodology is described in section~\ref{Methodology}.} 
 \label{fig:pairwise_delta_R2}
\end{figure}

We now focus on the pair of assets with tenors $\theta$ and $\theta'$. Our aim is to measure the degree to which the goodness-of-fit on the forward rate $\theta$ in a linear cross-impact model results from the order flow at $\theta'$. For this purpose, we define the accuracy increase of cross-sectional information as 
\begin{equation}
    \Delta \mathcal{R}^{2,\text{model}}_{\theta'\to \theta} := \mathcal{R}^{2,\text{model}}(W_{\sigma_\theta}) - \mathcal{R}^{2,\text{diag}}(W_{\sigma_\theta}),
\end{equation}
where the R-squared values are computed in a two-asset model. \citet{LeCozEtAl-2024a} established that the price-volume correlation between different US sovereign bonds depends on the respective liquidity of the considered assets. Formally \citet{LeCozEtAl-2024a} derive these results by observing that the pairwise additional R squared $\Delta \mathcal{R}^{2}_{\theta'\to \theta}$ obtained by the regression of a bond price on the order flow of another bond is highly asymmetrical. We reproduce these results for SOFR Futures contracts in Fig.~\ref{fig:pairwise_delta_R2}. The vertical stripes show the effect of the liquidity of each asset on the price-volume correlation. In fact, in a $2$-asset framework, the additional R-squared $\Delta \mathcal{R}^{2}_{\theta'\to \theta}$ is roughly equivalent to the squared price-volume correlation $\rho^2(\Delta f_\theta, \Delta q_{\theta'})$, due to the low spatial correlation of the order flow (see Fig.~\ref{fig:3dplot_correlation_volumes_screen_shot} and section~\ref{Maximum likelihood model}).

In this section, we demonstrate that the BBDLW and BBDLS models capture this stylized fact, primarily due to the vector $Y$, which represents the share of price volatility attributable to trades. These liquidity-dependent responses are neither replicated by the Kyle model (see section~\ref{Kyle model}) nor by the noise fields $A$ and $\eta$ (see section~\ref{Modeled responses}), highlighting the significance of the parameter vector $Y$.

\subsection{Theoretical ML model} \label{Maximum likelihood model}
In a two-asset ML model, the R-squared obtained from regressing the prices of the first asset to the order flow imbalance of the both assets is given by
\begin{align}
    &\mathcal{R}^{2,\text{ML}}(W_{\sigma_1})=\nonumber \\
    &\hspace{0.5cm}\frac{1}{1-\rho^2(\Delta q_1, \Delta q_2)} \left[ \rho^2(\Delta f_1, \Delta q_1) + \rho^2(\Delta f_1, \Delta q_2) \right. \nonumber \\
    & \hspace{0.8cm}\left.- 2 \rho(\Delta q_1, \Delta q_2) \rho(\Delta f_1, \Delta q_1) \rho(\Delta f_1, \Delta q_2)\right],
\end{align}
where $\Delta p_i$ and $\Delta q_i$ are respectively the price increments and the order flow  of assets $i \in \llbracket1,2\rrbracket$. If one subtracts from the previous quantity the R-squared obtained when regressing the first asset prices on its own trading flow, one gets the theoretical added accuracy $\Delta \mathcal{R}^{2,\text{ML}}_{2\to 1}$ in the ML model when regressing asset $1$'s prices on the order flow imbalance of assets $1$ and $2$ instead of solely asset $2$:
\begin{align}
    \label{eq:theo_R2_ml}
    \Delta \mathcal{R}^{2,\text{ML}}_{2\to 1} &=\mathcal{R}^{2,\text{ML}}(W_{\sigma_1}) - \rho^2(\Delta f_1, \Delta q_1) \nonumber \\
    &=\frac{(\rho(\Delta f_1, \Delta q_2) - \rho(\Delta q_1, \Delta q_2) \rho(\Delta f_1, \Delta q_1))^2}{1-\rho^2(\Delta q_1, \Delta q_2)}.
\end{align}
Equation~\eqref{eq:theo_R2_ml} indicates that $\Delta \mathcal{R}^{2,\text{ML}}_{2\to 1}$ depends solely on price and order flow correlations. A priori $\Delta \mathcal{R}^{2,\text{ML}}_{2\to 1}$ should be independent of the respective liquidity (i.e., the product $\sigma_i \times \omega_i$, where $\sigma_i$ is the volatility of prices and $\omega_i$ the volatility of the order flow imbalance) of each asset. In fact, Fig.~\ref{fig:3dplot_correlation_volumes_screen_shot} illustrates that these R-squared values vary significantly across assets, suggesting that price-volume correlations are influenced by the liquidity of the assets in question (see \citet{LeCozEtAl-2024a} for a detailed analysis of liquidity's effect on cross-impact).

\subsection{Theoretical Kyle model} \label{Kyle model}
\begin{figure}
\centering
 \includegraphics[width=0.9\linewidth]{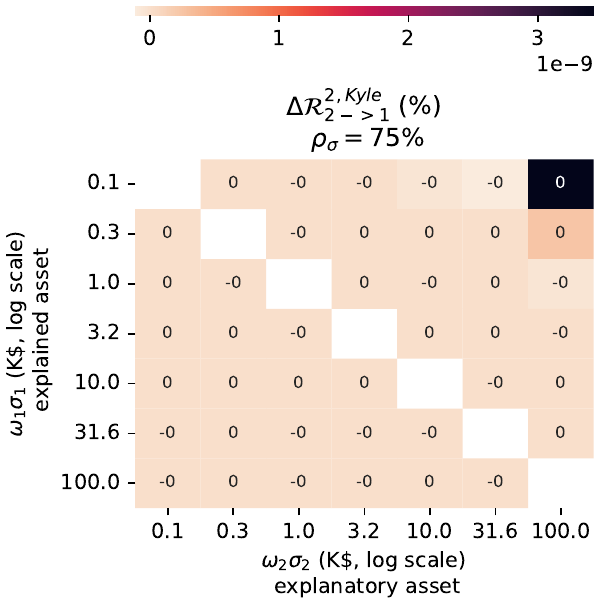}
 \caption{Theoretical added accuracy $\Delta \mathcal{R}^{2}_{2\to 1}$ in the Kyle model, when regressing the price of asset $1$ on the order flow imbalance of assets $1$ and $2$ instead of solely asset $2$. $\Delta \mathcal{R}^{2,\text{Kyle}}_{2\to 1}$ is represented as a function of the individual risk levels of each asset. The correlation between the order flows of assets 1 and 2 is $ \rho_{\omega} = 50\%$. The correlation between the prices of assets 1 and 2 is $ \rho_{\omega} = 75\%$. The volatility of prices and volumes is defined as the square root of the risk level: $\sigma_1 = \omega_1 = \sqrt{\sigma_1 \omega_1}$.} 
 \label{fig:theo_kyle}
\end{figure}

A numerical simulation clearly demonstrates the effect of liquidity in the Kyle model. Figure~\ref{fig:theo_kyle} shows the added precision $\Delta \mathcal{R}^{2,\text{Kyle}}_{2\to 1}$ in the Kyle model, when regressing the prices of the asset $1$ on the order flow imbalance of the assets $1$ and $2$ instead of solely on the asset $2$. Each pair of assets is identified by the liquidity $\sigma_i \omega_i$ of its individual assets. We assume that the $Y$-ratio remains constant for all pairs considered and $y = \rho(\Delta f_1, \Delta q_1)$. In other words, the $Y$-ratio is precisely equal to the correlation between price and volume for the explained asset. Figure~\ref{fig:theo_kyle} shows that, in this scenario, $\Delta \mathcal{R}^{2,\text{Kyle}}_{2\to 1}$ is close to zero for all the liquidity levels tested. This means that the Kyle model consistently generates an R-squared $\mathcal{R}^{2,\text{Kyle}}(W_\sigma)$ close to $y^2$. It yields another interpretation of the Y-ratio in the Kyle model as the average effective correlation between prices and volumes. 

\subsection{Theoretical BBDL models} \label{Modeled responses}

\begin{figure}
\centering
 \includegraphics[width=\linewidth]{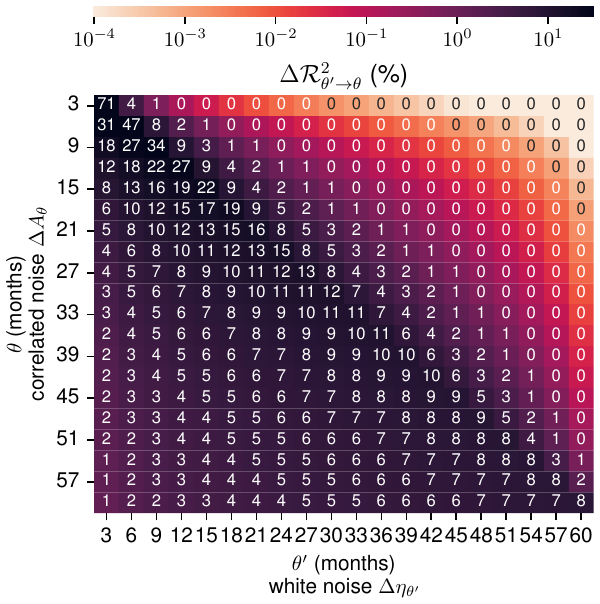}
 \caption{Theoretical additional R-squared from regressing $\Delta A_{\theta}(t)$ on $\Delta \eta_{\theta'}(t)$ and $\Delta \eta_{\theta}(t)$ instead of solely $\Delta \eta_{\theta}(t)$, according to Eq.~\eqref{eq:corr_DeltaA_Deltaeta}. The parameter $\kappa$ is calibrated on forward rate correlations for the period $2021-2023$ (i.e., $\kappa=1.3$).}
 \label{fig:pairwise_theo_BB}
\end{figure}

\paragraph{Responses of $A$ to $\eta$.} We first study the response of the correlated field $A$ to its generating white noise $\eta$ given by Eq.~\eqref{eq:corr_DeltaA_Deltaeta}. Figure~\ref{fig:pairwise_theo_BB} shows the squared correlation between $\Delta A_{\theta}(t)$ and $\Delta \eta_{\theta'}(t)$ for a typical value of the calibrated parameter~$\kappa$. This quantity represents the additional R squared from the regression of $\Delta A_{\theta}(t)$ to $\Delta \eta_{\theta'}(t)$ and $\Delta \eta_{\theta}(t)$ instead of only $\Delta \eta_{\theta}(t)$. Indeed, as $\Delta \eta_{\theta}(t)$ in independent from $\Delta \eta_{\theta'}(t)$ we have
\begin{align}
    \Delta \mathcal{R}^{2}_{\theta'\to \theta} = \frac{\E{\Delta A_{\theta}(t) \Delta \eta_{\theta'}(t)}^2}{\E{\Delta A_{\theta}(t)^2} \E{\Delta \eta_{\theta'}(t)^2}} =\frac{R^2_{\theta\theta'}}{(\sigma_A)_\theta}.
\end{align}
Figure~\ref{fig:pairwise_theo_BB} shows that the correlation between noise $A$ and its generating white noise $\eta$ is asymmetrical and decreases with distance $|\theta-\theta'|$. This asymmetry arises from the rescaling by the norm $\sigma_A$ of the noise field $A$, which decreases when $\theta$ increases. This decreasing volatility is an effect of psychological time: the higher the maturity, the shorter the distance between the nearby tenors; thus, the lower the volatility of each noise $A_{\theta'}$ generated from the normalized white noise $\eta_{\theta}$.

\begin{figure}
\centering
 \includegraphics[width=\linewidth]{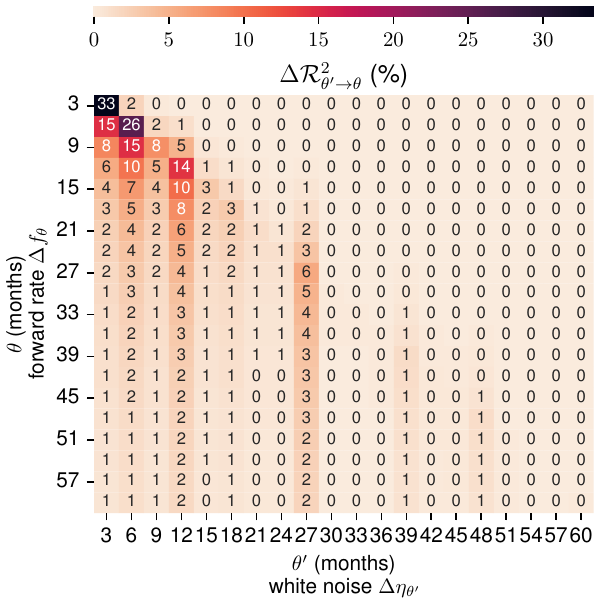}
 \caption{Theoretical additional R-squared from regressing $\Delta f_{\theta}(t)$ on $\Delta \eta_{\theta'}^q(t)$ and $\Delta \eta_{\theta}^q(t)$ instead of solely $\Delta \eta_{\theta}^q(t)$. The parameter $\kappa$ is calibrated on forward rate correlations for the period $2021-2023$ (i.e., $\kappa=1.3$) and $Y$ is calibrated on the same period using the BBDLW model.}
 \label{fig:pairwise_theo_BB_Y}
\end{figure}

\paragraph{Responses of the forward rate to $\eta$.} In the BBDLW model, the additional R-squared from regressing $\Delta f_{\theta}(t)$ on $\Delta \eta_{\theta'}^q(t)$ and $\Delta \eta_{\theta}^q(t)$ instead of solely $\Delta \eta_{\theta}^q(t)$ is given by
\begin{align}
    \label{eq:delta_R2_f_eta}
    \Delta \mathcal{R}^{2}_{\theta'\to \theta} = \frac{\E{\Delta f_{\theta}(t) \Delta \eta_{\theta'}(t)}^2}{\E{\Delta f_{\theta}(t)^2} \E{\Delta \eta_{\theta'}(t)^2}} = (\diag(\sigma_A)^{-1}R)^2_{\theta\theta'} Y_{\theta'}^2.
\end{align}
Figure~\ref{fig:pairwise_theo_BB_Y} shows that the correlation between the forward rate and its generating white noise $\eta^q$ exhibits vertical stripes related to the liquidity of the products considered. This is an effect of differences in the share $Y_\theta$ of the volatility explained by each white noise $\eta_\theta$. However, the model cannot correct for the decreasing volatility of the noise field $A$, as shown by the lower R-squared in the top right of the matrix in Fig.~\ref{fig:pairwise_theo_BB_Y}. We would have obtained similar results in the BBDLS model, although it would require inserting a rotation factor $O^2_{\text{sym}}(\diag(\sigma_A)^{-1} R \diag(Y), I)$ into Eq.~\eqref{eq:delta_R2_f_eta}. As shown in the following section, this rotation also creates horizontal stripes that correspond to the symmetrization of the cross-impact matrix.

\begin{figure}
\centering
 \includegraphics[width=\linewidth]{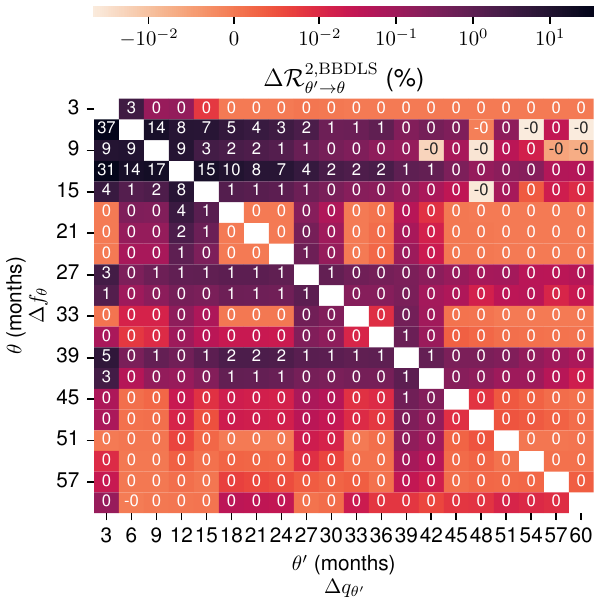}
 \caption{Theoretical additional R-squared from regressing $\Delta f_{\theta}(t)$ on $\Delta q_{\theta'}(t)$ and $\Delta q_{\theta}(t)$ instead of solely $\Delta q_{\theta}(t)$. The parameter $\kappa$ is calibrated on forward rate correlations for the period $2021-2023$ (i.e., $\kappa=1.3$) and $Y$ is calibrated on the same period using the BBDLS model.}
 \label{fig:pairwise_theo_BB_Y_rho}
\end{figure}

\paragraph{Responses of the forward rate to the order flow.} In the BBDLS model, the additional R-squared from regressing $\Delta f_{\theta}(t)$ on $\Delta q_{\theta'}(t)$ and $\Delta q_{\theta}(t)$ instead of solely $\Delta q_{\theta}(t)$ is given by
\begin{align}
    \label{eq:delta_R2_f_q}
    \Delta \mathcal{R}^{2,\text{BBDLS}}_{\theta'\to \theta} = \lambda_{\theta\theta'}^2\omega^2_{\theta'} + 2\rho^{q}_{\theta\theta'} \lambda_{\theta\theta}\lambda_{\theta\theta'} \omega_\theta \omega_{\theta'},
\end{align}
where $\lambda = \diag(\sigma_A)^{-1}R\diag(Y)O_{\text{sym}}\Omega^{-1/2}$ is $2\times2$ normalized cross-impact matrix in the BBDLS model. More precisely, the matrix $\diag(\sigma_A)^{-1}R\diag(Y)$ is given by the model of dimension $n$ restricted to Futures contracts of tenor $\theta$ and $\theta'$. The matrix $\Omega^{-1/2}$ is defined from a matrix $\Omega$ restricted to two Futures contracts of tenor $\theta$ and $\theta'$.

Figure~\ref{fig:pairwise_theo_BB_Y_rho} shows that these additional R-squares exhibit vertical stripes related to the liquidity of the products considered. Although we reproduce the order of magnitude of the empirical measures (see Fig.~\ref{fig:pairwise_delta_R2}), we do not precisely match the observed R-squared. In fact, we only have one parameter $Y_{\theta'}$ per column to correct the asymmetric shape of the correlation between $\Delta A$ and $\Delta \eta$ in Fig.~\ref{fig:pairwise_theo_BB}. Thus, while the price-volume correlation depends on the respective liquidity of the considered asset pair \citep{LeCozEtAl-2024a}, in our model, it depends only on the explanatory asset. For a given order flow $\theta$, we optimize the average $Y_\theta$ that best matches the liquidity of all assets.

Furthermore, in contrast with the empirical results, we observe that the BBDLS model generates horizontal stripes in the additional pairwise R-squared values (see Fig.~\ref{fig:pairwise_theo_BB_Y_rho}). This occurs because of the symmetry of the cross-impact matrix imposed by the absence of arbitrage. As mentioned previously, the spatial correlation of the order flow is low (see Fig.~\ref{fig:3dplot_correlation_volumes_screen_shot}), rendering the second term in Eq.~\eqref{eq:delta_R2_f_q} negligible. Consequently, pairwise R-squared values are primarily influenced by the term $\lambda_{\theta\theta'}^2\omega^2_{\theta'}$. We have seen that in the Kyle model, the liquidity $\omega^2_{\theta'}$ does not alter the price-volume correlation, so the R-squared values in Fig.~\ref{fig:pairwise_theo_BB_Y_rho} can be approximated by the product $YY^\top$, which is symmetric.

\section{Conclusion}

Let us start by summarizing what we have achieved. First, we have shown that the BBDL model \citep{LeCozBouchaud-2024a} is consistent with the well-documented temporal autocorrelation of forward rates at short time scales. It appears that the time scale $\tau$ at which spatial correlations among asset prices emerge \citep{Epps-1979} is also the point where temporal correlations begin to dissipate. Therefore, this framework describes how the spatial and temporal correlation structure of prices evolves across time scales.

Most importantly, we have proposed a new interpretation of the BBDL model \citep{LeCozBouchaud-2024a} in which high-frequency shocks are identified to trades. The latter, which exhibit low spatial correlations, affect each point of the interest rate curve independently on the smallest time scale. The spatial correlation structure of prices emerges from market participants reacting to these independent trades and external shocks, such as news events, that simultaneously affect multiple points along the yield curve. These participants then propagate the impact of these shocks across other maturities through a self-referential mechanism, as described by \citet{LeCozBouchaud-2024a}.

Consequently, this model can be interpreted as a cross-impact model, linking order flows to price movements. A key feature is that only the surprise component of trades influences prices, similarly to a propagator model. To address the challenge of temporal independence, we calibrate the model at a daily time scale, where trades show low autocorrelation. Using this approach, we can match or exceed the precision of the multivariate Kyle model in fitting price moves to order flows, but with far fewer parameters ($n+1$ compared to $\frac{n(n-1)}{2}$). Furthermore, unlike the Kyle model, this framework accounts for liquidity-dependent correlations between the forward rate of one maturity and the order flow of another.

A promising direction for future research is to explore the micro-level mechanisms that connect liquidity with price-volume correlations. This would likely involve the development of a multidimensional model of the limit order book, shedding light on liquidity dynamics across different assets and maturities.

\section{Acknowledgments}
We express our gratitude to Jean-Philippe Bouchaud and Damien Challet, who contributed to our research through fruitful discussions. We also thank Bertrand Hassani and the ANRT (CIFRE number 2021/0902) for providing us with the opportunity to conduct this research at Quant AI Lab. We extend our appreciation to Medhi Tomas for his assistance in designing the temporal and spatial decomposition of the order flow. 

This research was carried out within the Econophysics \& Complex Systems Research Chair, under the aegis of the Fondation du Risque, the Fondation de l’École polytechnique, the École polytechnique, and Capital Fund Management.

\bibliographystyle{apsrev4-2} 
\bibliography{zotero}

\newpage
\appendix
\section{Notations} \label{Notations}
Table~\ref{tab:notations} summaries the notations used in this study. 
\begingroup
\squeezetable
\begin{table}[h]
    \begin{ruledtabular}
    \begin{tabular}{p{0.1\linewidth} p{0.85\linewidth}}
    \multicolumn{1}{c}{Expression} & \multicolumn{1}{c}{Definition} \\
    \hline
    $n$ & The number of available SOFR Futures. \\
    $\mathbf{M}_n(\mathbb{R})$ & The set of real-valued square matrices of dimension $n$.\\
    $M^\top$ & The transpose of matrix~$M$. \\
    $\diag(M) $ & The vector in $\mathbb{R}^n$ formed by the diagonal items of the matrix $M$. \\
    $\diag(v)$ & The diagonal matrix whose components are the components $(v_1, \cdots, v_n)$ of the vector $v \in \mathbb{R}^n$.\\
    $M^{1/2}$ & A matrix such that $M^{1/2}(M^{1/2})^\top = M$. \\
    $\sqrt{M}$ & The unique positive semi-definite symmetric matrix such that $(\sqrt{M})^2 = M$. \\
    $\Lambda(t)$ &  The cross-impact matrix at time $t$. \\
    $\sigma(t)$ & The vector of price variation volatility at time $t$. \\
    $\omega(t)$ & The vector of the signed order flow volatility at time $t$. \\
    $\mathcal{R}^2(W)$ & The $W$-weighted generalized R-squared. \\
    $\Delta \mathcal{R}^2(W)$ & The accuracy increase from the cross sectional model. \\
    $t$ & The current time. \\
    $T$ & The maturity.\\
    $P(t,T)$ & The price at time t of a zero-coupon bond maturing at~$T$. \\
    $\theta$ & The time-to-maturity or tenor. \\
    $f_\theta(t)$ & The value at time t of the instantaneous forward rate of tenor~$\theta$ (discrete notation). \\
    $f(t)$ & The vector of forward rates at time~$t$. \\
    $\Delta q_\theta(t)$ & The net market order flow traded during the time window $[t, t + \Delta t]$. \\
    $\Delta q(t)$ & The vector of the net traded order flows during the time window $[t, t + \Delta t]$. \\
    $A_\theta(t)$ & The driftless correlated noise field. \\
    $\eta_{\theta}(.)$ & The discrete white noise of tenor~$\theta$. \\
    $\sigma_\theta(t)$ & The volatility at time~$t$ of the infinitesimal variation of the instantaneous forward rate of time-to-maturity~$\theta$. \\
    $\mu$ & The line tension parameter. \\
    $\psi$ & The psychological time parameter. \\
    $\kappa$ & Unique a-dimensional parameter in the BBDL model, defined as the product $\mu \times\psi$. \\
    $\tau$ & The time scale for the emergence of correlations. \\
    $\Delta t$ & The temporal duration of a day. \\
    $\E{.}$ & The unconditional expectancy. \\
    $\langle . \rangle(t)$ & The empirical average operator over the interval $[t-\Delta t, t]$. \\
    $\hat{x}(t)$ & The estimator of~$x$ at time~$t$. \\
    $\eta_{\theta}(.)$ & The discrete white noise of tenor~$\theta$. \\
    $\delta(.)$ & The Dirac delta function. \\
    $\rho(x,y)$ & The linear Pearson correlation matrix between the random vector $x$ and $y$. \\
    $\delta_{\theta\theta'}$ & The Kronecker delta. \\
    $I_k$ & A matrix with ones only on the $k$-th diagonal above the main diagonal. \\
    $I$ & The identity matrix. \\
    $\mathcal{J}$ & A diagonal matrix whose first entry is $2$ while all the other entries are ones. \\
    $\mathcal{L}_d[.] $ & The discrete linear differential operator on space. \\
    $\mathcal{M} $ & The discrete non-linear differential operator on space, using matrix notations. \\
    $L_d[.] $ & The Fourier transform of the discrete linear differential operator on space. \\
    $G_{\theta\theta'}(.) $ & Green function or \textit{propagator} of Eq.~\eqref{eq:discret_master_general} \\
    $N$ & The number of days in a $3$-year period of our sample. \\
    $H(.)$ & The Heaviside function. \\
    $\mathcal{F}[f]$ & The Fourier transform of the function of time~$f$.
    \end{tabular}
    \end{ruledtabular}
    \caption{Notations\label{tab:notations}}
\end{table}
\endgroup

\FloatBarrier
\section{Responses to the white noise for $\psi \gg 1$} \label{Response to the white noise for psi above 1}
\subsection{Covariance}
We derive the covariance between $\Delta A_{\theta}(t)$ and $\Delta \eta_{\theta'}(t)$ in the limit $\psi \gg 1$:
\begin{align}
    \label{eq:ap_cov_A_eta_large_psi}
    & \E{\Delta A_{\theta}(t) \Delta \eta_{\theta'}(t)} \nonumber\\
    &=  \int_{t-\Delta t}^t {\rm d}u \int_{t-\Delta t}^{t} {\rm d}v \E{A_{\theta}(u) \eta_{\theta'}(v) } \nonumber\\
    &= \frac{1}{\tau} \int_{t-\Delta t}^t {\rm d}u \int_{t-\Delta t}^{u} {\rm d}v G_{\theta,\theta'}(u-v) \nonumber\\
    &= \frac{1}{2\pi\tau}\int_{-\pi}^{\pi}{\rm d}\xi \left( e^{i\xi(\theta-\theta')} + e^{i\xi(\theta+\theta')} \right) \nonumber\\
    & \hspace{3cm} \int_{t-\Delta t}^t {\rm d}u \int_{t'-\Delta t}^{u} {\rm d}v e^{-\frac{L_{d}(\xi)}{\tau}(u-v)} \nonumber\\
    &= \frac{1}{2\pi} \int_{-\pi}^{\pi}{\rm d}\xi \frac{e^{i\xi(\theta-\theta')} + e^{i\xi(\theta+\theta')}}{L_{d}(\xi)} \nonumber\\
    & \hspace{3cm} \int_{t-\Delta t}^t {\rm d}u \left(1-e^{-\frac{L_{d}(\xi)}{\tau}(u-t+\Delta t)}\right) \nonumber \\
    & \xrightarrow[\tau \mapsto 0]{}   \Delta t (\mathcal{D}_1)_{\theta\theta'}.
\end{align}

\subsection{Correlation}
Using Eq.~\eqref{eq:ap_cov_A_eta_large_psi}, we obtain the correlation between $\Delta A_{\theta}(t)$ and $\Delta \eta_{\theta'}(t)$:
\begin{align}
    \label{eq:correl_A_eta}
    & \frac{\E{\Delta A_{\theta}(t) \Delta \eta_{\theta'}(t)}}{\sqrt{\E{ \Delta A_{\theta}(t)^2 } \E{\Delta \eta_{\theta'}(t)^2}}} = \frac{(\mathcal{D}_1)_{\theta\theta'}}{\sqrt{(\mathcal{D}_2)_{\theta\theta}}}.
\end{align}

We can show that Eq.~\eqref{eq:correl_A_eta} is well defined. For this purpose, we define the usual inner product $(f,g)$ between two integrable real-valued functions $f$ and $g$ on $[0,\pi]$ by:
\begin{align}
    \label{eq_inner_product}
    (f,g) = \frac{1}{\pi}\int_{0}^{\pi}{\rm d}\xi f(\xi)g(\xi).
\end{align}
For $(\theta,\theta') \in \llbracket 1,n\rrbracket^2$, having noted that $\frac{2}{\pi}\int_{0}^{\pi}{\rm d}\xi\cos^2{\xi \theta}=1$, we have:
\begin{align}
    & (\mathcal{D}_1)_{\theta\theta'} = \left(\frac{\sqrt{2}\cos{\xi \theta}}{L_{d}(\xi)},\sqrt{2}\cos{\xi \theta'}\right), \\
    &  (\mathcal{D}_2)_{\theta\theta} = \left(\frac{\sqrt{2}\cos{\xi \theta}}{L_{d}(\xi)},\frac{\sqrt{2}\cos{\xi \theta}}{L_{d}(\xi)}\right) \Bigg(\sqrt{2}\cos{\xi \theta'},\sqrt{2}\cos{\xi \theta'}\Bigg).
\end{align}
Thus, Cauchy-Schwarz's inequality ensures that 
\begin{align}
   -1\leq \frac{(\mathcal{D}_1)_{\theta\theta'}}{\sqrt{(\mathcal{D}_2)_{\theta\theta}}} \leq 1.
\end{align}

\section{Responses to the white noise for $\psi \ll 1$} \label{Response to the white noise for psi below 1}
\subsection{Covariance}
We derive the covariance between $\Delta A_{\theta}(t)$ and $\Delta \eta_{\theta'}(t)$ in the limit $\psi \ll 1$:
\begin{align}
    \label{eq:ap_cov_A_eta_small_psi}
    & \E{\Delta A(t) \Delta \eta(t)^\top } \nonumber\\
    &=  \int_{t-\Delta t}^t {\rm d}u \int_{t-\Delta t}^{t} {\rm d}v \E{A(u) \eta(v)^\top } \nonumber\\
    &= \frac{1}{\tau} \int_{t-\Delta t}^t {\rm d}u \int_{t-\Delta t}^{u} {\rm d}v e^{-\frac{u-v}{\tau}\mathcal{M}} \mathcal{J}\nonumber\\
    &= \int_{t-\Delta t}^t {\rm d}u \mathcal{M}^{-1}\left(1-e^{-\frac{ u-t+\Delta t}{\tau}\mathcal{M}} \right)\mathcal{J} \nonumber \\
    & \xrightarrow[\tau \mapsto 0]{}   \Delta t \mathcal{M}^{-1} J.
\end{align}

\subsection{Correlation}
Using Eq.~\eqref{eq:ap_cov_A_eta_small_psi}, we obtain the correlation between $\Delta A_{\theta}(t)$ and $\Delta \eta_{\theta'}(t)$:
\begin{align}
    & \frac{\E{\Delta A_{\theta}(t) \Delta \eta_{\theta'}(t) }}{\sqrt{\E{\Delta A_{\theta}(t)^2 } \E{\Delta \eta_{\theta'}(t)^2}}} = \frac{\left(\mathcal{M}^{-1}\mathcal{J}\right)_{\theta\theta'}}{\sqrt{\left(\mathcal{M}^{-1}\mathcal{J}^2(\mathcal{M}^{-1})^\top\right)_{\theta\theta}}},
\end{align}
which is clearly well defined.

\section{Order flow decomposition} \label{app:Order flow decomposition}
Let $\eta^q$ be a white noise vector such that $\E{\eta^{q}_{\theta}(t)\eta^{q}_{\theta'}(t')} = \delta_{\theta\theta'} \delta(t-t')$. We decompose the trading flow across time and space into the cumulative sum of~$\eta^q$ and a kernel~$K$:
\begin{align}
    \label{eq:flow_decomposition_app}
    \frac{{\rm d}q}{{\rm d}t}(t) = \int_{-\infty}^{t} {\rm d}t' K(t,t') O^{\top} \eta^{q}(t'),
\end{align}
where $O$ is an orthogonal matrix and the kernel $K(u,u') \in \mathbf{M}_n(\mathbb{R})$ is such that the lagged variance-covariance matrix $\Omega(t,t')$ of the infinitesimal order flow imbalance reads
\begin{align}
    \Omega(t,t') :=& \E{\frac{{\rm d}q}{{\rm d}t}(t) \frac{{\rm d}q}{{\rm d}t}^{\top}(t')} \nonumber \\
    =&\int_{-\infty}^{t} {\rm d}u \int_{-\infty}^{t'} {\rm d}v K(t,u) O^{\top} \E{\eta^q(u) \eta^q(v)^{\top}} \nonumber \\
    & \hspace{5cm} O K^{\top}(t',v)  \nonumber \\
    =& \int_{-\infty}^{\min(t,t')} {\rm d}u K(t,u) {K}^{\top}(t',u).
\end{align}
We assume that the series of matrices $(K(t,t'))_{t\ge0,t'\ge0}$ can be written $K(t-t')_{t-t'\ge0}$. Under this assumption, the lagged variance-covariance matrix $\Omega(t,t')$ is stationary and can be written as a function of the lag $\ell = t-t'$:
\begin{align}
    \label{eq:autocorrelation}
     \Omega(t,t-l) &= \int_{\mathbb{R}}{\rm d}u K(t-u) K^{\top}(t-\ell-u) \nonumber \\
     &= \int_{\mathbb{R}}{\rm d}u' K(u') K^{\top}(u'-\ell).
\end{align}
The Fourier transform over the lag $\ell$ of Eq.~\eqref{eq:autocorrelation} reads
\begin{align}
    \mathcal{F}[\Omega](m) = \mathcal{F}[K](m) \{\mathcal{F}[K](m)\}^{\top},
\end{align}
where $\mathcal{F}[K]$ and $\mathcal{F}[\Omega]$ are the Fourier transform of $\ell \to K(\ell)$ and $\ell \to \Omega(\ell)$. Thus, Eq.~\eqref{eq:flow_decomposition_app} is verified if and only if, each matrix $\mathcal{F}[K](m)$ is a decomposition of $\mathcal{F}[\Omega](m)$:
\begin{align}
    \mathcal{F}[K](m)= \{\mathcal{F}[\Omega](m)\}^{1/2}.
\end{align}
For example, we can build numerically each $\mathcal{F}[K](m)$ as the Cholesky decomposition of $\mathcal{F}[\Omega](m)$.

Still assuming stationarity, the definition of $q(t)$ in Eq.~\eqref{eq:flow_decomposition_app} reads as the convolution product of $K$ and $\eta$. The Fourier transform over time of Eq.~\eqref{eq:flow_decomposition_app} reads in matrix notations
\begin{align}
    \label{eq:fourier_convolution}
    \mathcal{F}\left[\frac{{\rm d}q}{{\rm d}t}\right](m) = \mathcal{F}[K](m)O^{\top}\mathcal{F}[\eta^{q}](m).
\end{align}
Assuming the matrix $\hat{K}(m)$ is invertible one can write Eq.~\eqref{eq:fourier_convolution} as 
\begin{align}
    \mathcal{F}[\eta^{q}](m) = O\{\mathcal{F}[K](m)\}^{-1}\mathcal{F}\left[\frac{{\rm d}q}{{\rm d}t}\right](m).
\end{align}
Hence, one can also write the white noise~$\eta^q$ as the convolution product: 
\begin{align}
    \label{eq:white_noise_building_app}
    \eta^{q}(t) &= O \int_{-\infty}^t{\rm d}t' J(t-t') \frac{{\rm d}q}{{\rm d}t}(t'), \nonumber \\
\end{align}
where the function $u \mapsto J(u)$, valued in $\mathbf{M}_n(\mathbb{R})$, is the inverse Fourier transform of $m \mapsto \{\mathcal{F}[K](m)\}^{-1}$.

\section{Large-bin approximation} \label{appendix large-bin approximation}
In this section, we denote by $\bar{x}_{\Delta t}$ the observed empirical average over the time interval $[t - \Delta t,t]$ of the random process $x(t)$ (i.e., its moving average):
\begin{align}
    \bar{x}_{\Delta t}(t) &:= \frac{1}{\Delta t}\int_{t-\Delta t}^t {\rm d}t' x(t').  
\end{align}
We aim to approximate $\Delta A_\theta(t)$ as a function of coarse-grained variables $\Delta \eta$ defined over intervals of finite width $\Delta t$. For this purpose, we decompose the white noise $\eta(t')$ as the sum of its moving average and its fluctuations around this mean. Formally, we write
\begin{align}
    \label{eq:eta_decomposition_appendix}
    \eta(t') = \bar{\eta}_{\Delta t}(t) + \eta(t') - \bar{\eta}_{\Delta t}(t).
\end{align}
The independence and stationarity of $\eta$ across time ensures that $\bar{\eta}_{\Delta t}(t)$ is uncorrelated with $\eta(t') - \bar{\eta}_{\Delta t}(t)$:
\begin{align}
    &\E{\bar{\eta}_{\Delta t}(t) \left( \eta(t') - \bar{\eta}_{\Delta t}(t)\right)} \nonumber \\
    &= \frac{1}{\Delta t} \int_{t-\Delta t}^t {\rm d}t'' \E{\eta(t')\eta(t'')} \nonumber \\
    &\hspace{1cm}- \frac{1}{(\Delta t)^2} \iint_{t-\Delta t}^t {\rm d}t'' {\rm d}t''' \E{\eta(t'')\eta(t''')} \nonumber \\
    &= \frac{1}{\Delta t} - \frac{1}{\Delta t} = 0.
\end{align}
In fact, the continuous-time hypothesis, which is related to the Gaussianity of the Langevin noise $\eta$, ensures the independence between $\bar{\eta}_{\Delta t}(t)$ and $\eta(t') - \bar{\eta}_{\Delta t}(t)$ (although only an absence of correlation is needed here). It is worth mentioning that this result can also be derive by observing that the process $\int_{0}^t {\rm d}t'\eta(t') - \int_{0}^t {\rm d}t'\bar{\eta}_{\Delta t}(t')$ is a Brownian bridge.

We now define the matrix $R^{\tau}(t)$ by
\begin{align}
    R^{\tau}(t) = \frac{1}{\tau} \int_{-\infty}^t {\rm d}t' G(t-t').
\end{align}
Correlations among assets appear at a time scale $\tau \ll \Delta t $ \citep{Epps-1979,LeCozBouchaud-2024a} and the time decay of $G$ is very strong : $G_{\theta\theta'}(5\tau)/G_{\theta\theta'}(0) \approx 10^{-3}$ for typical values of $\kappa \approx 1$. Hence, we can approximate $R^{\tau}(t)$ by
\begin{align}
    \label{eq:M_tau_approx}
    R^{\tau}(t) \approx \frac{1}{\tau} \int_{t-\Delta t}^t {\rm d}t' G(t-t').
\end{align}
Substituting $\eta$ by Eq.~\eqref{eq:eta_decomposition_appendix} in the definition~\eqref{eq:corr_noise_def} of the noise field $A(t)$ yields
\begin{align}
    \label{eq:A_lin}
    A(t) = R^{\tau}(t) \; \bar{\eta}_{\Delta t}(t) + \epsilon^{\tau}(t),
\end{align}
where $\epsilon^{\tau}(t) = \frac{1}{\tau} \int_{t-\Delta t}^t {\rm d}t' G(t-t') \left( \eta(t') - \bar{\eta}_{\Delta t}(t) \right)$ is a noise independent from $\bar{\eta}_{\Delta t}(t)$. Importantly, $\epsilon^{\tau}$ has no temporal correlation but has a spatial structure allowing to retrieve the spatial correlations of $A(t)$. The integration of Eq.~\eqref{eq:A_lin} over the interval $[t - \Delta t,t]$ yields an affine relationship between $\Delta A$ the sum of $A$ over one day, and the empirical daily means $\bar{\eta}_{\Delta t}(t)$:
\begin{align}
    \label{eq:delta_A_lin}
    \Delta A(t) = \left(\int_{t-\Delta t}^{t} {\rm d}t' R^{\tau}(t') \right)\bar{\eta}_{\Delta t}(t) + \int_{t-\Delta t}^{t} {\rm d}t' \epsilon^{\tau}(t'). 
\end{align}
One can substitute $\eta$ with $\eta^q$ or $\eta^{\top}$ and $A$ with $A^q$ or $A^{\top}$ respectively in Eq.~\eqref{eq:delta_A_lin}. Thus, having noted that $\langle \Delta A^{\perp}(t) \Delta \eta^q(t) \rangle = 0$, one can relate forward rate daily increments to the empirical daily means of the martingale component of the order flow:
\begin{align}
    \label{eq:delta_forward_lin}
    &\Delta f(t) = \nonumber \\
    & \diag(\sigma) \diag(\sigma_A)^{-1} \left( \int_{t-\Delta t}^{t} {\rm d}t' R^{\tau}(t') \right) \nonumber \\
    & \hspace{2cm}\diag(Y) O \Omega^{-1/2} \overline{\frac{{\rm d}\tilde{q}}{{\rm d}t}}_{\Delta t} (t)  +  \mathcal{E}^{\tau}(t),
\end{align}
where the residual noise $\mathcal{E}^{\tau}(t)$ is independent from $\overline{\frac{{\rm d}\tilde{q}}{{\rm d}t}}_{\Delta t} (t)$. Indeed, $\mathcal{E}^{\tau}(t)$ reads
\begin{align}
    &\mathcal{E}^{\tau}(t) = \nonumber \\
    &\diag(\sigma) \diag(\sigma_A)^{-1} \left( \int_{t-\Delta t}^{t} {\rm d}t' R^{\tau}(t') \right) \diag(Y^{\perp})  \overline{\eta^{\perp}}_{\Delta t}(t) \nonumber\\  
    & \hspace{1cm} + \diag(\sigma_A)^{-1} \int_{t-\Delta t}^{t} {\rm d}t' \epsilon^{\tau}(t').
\end{align}
Moreover, by definition of the empirical mean, we have
\begin{align}
    \overline{\frac{{\rm d}\tilde{q}}{{\rm d}t}}_{\Delta t} (t) = \frac{1}{\Delta t}\int_{t-\Delta t}^{t} {\rm d}t' \frac{{\rm d}\tilde{q}}{{\rm d}t'} (t') = \frac{\Delta \tilde{q} (t)}{\Delta t}.
\end{align}
We denote~$\widehat{\Delta f}$ the conditional expectancy of the forward rates increments~$\Delta f$ with respect to the martingale component of the order flows~$\Delta \tilde{q}(t)$: 
\begin{align}
    \widehat{\Delta f}(t) := \mathbb{E}\left[ \Delta f(t) \right| \left. \Delta \tilde{q}(t) \right].
\end{align}
Taking the conditional expectancy of Eq.~\eqref{eq:delta_forward_lin} yields
\begin{align}
    \label{eq:forward_hat}
   &\widehat{\Delta f}(t) =  \nonumber\\
   &\diag(\sigma) \diag(\sigma_A)^{-1} \left( \int_{t-\Delta t}^{t} {\rm d}t' R^{\tau}(t') \right) \diag(Y) O \Omega^{-1/2} \frac{\Delta \tilde{q}(t)}{\Delta t}.
\end{align}
One can choose $\tau$ arbitrarily small in the expression of $R^{\tau}(t)$. In the limit $\tau \ll 1$ we have
\begin{align}
    R^{\tau}(t) &\xrightarrow[\tau \to 0]{} R.
\end{align}
Thus, in this limit, Eq.~\eqref{eq:forward_hat} reads
\begin{align}
     \widehat{\Delta f}(t) & = \diag(\sigma) \diag(\sigma_A)^{-1} R \diag(Y) O \Omega^{-1/2}\Delta \tilde{q}(t).
\end{align}
In this model, the residual noise $\mathcal{E}$ in Eq.~\eqref{eq:definition_lambda} can be seen as the limit for small $\tau$ of $\mathcal{E}^{\tau}(t)$.

\section{Response to order flows} \label{Response to order flows for psi above 1}
In this appendix we show that 
\begin{align}
    &\E{\Delta f(t) \Delta q^\top(t)} \E{\Delta q(t) \Delta q^\top(t)}^{-1} \nonumber \\
    & \hspace{1cm}  =\diag(\sigma) \diag(\sigma_A)^{-1} R \diag(Y) O \Omega^{-1/2}.
\end{align}
For this purpose we first derive the expression of the covariance $\E{\Delta f(t) \Delta q(t)^\top}$. We also show that the correlation between the forward rates and the order flows is well defined.

\subsection{Computation of the covariance matrix}
In this section we derive the expression of the covariance between forward rates and order flows.

We define the accumulated trading flows over the period $[t,t+\Delta t]$, representing a trading day, as
\begin{align}
    \Delta q(t) = \int_{t-\Delta t}^{t} {\rm d}t' \frac{{\rm d}q}{{\rm d}t}(t').
\end{align}
The equal-time covariance between $\Delta A^q(t)$ and $\Delta q(t)$ is
\begin{align}
    \E{\Delta A^q(t) \Delta q^\top(t) } =  \iint_{t-\Delta t}^t {\rm d}u{\rm d}v \int_{-\infty}^v {\rm d}v' \hspace{1cm}\nonumber \\
    \E{A^q(u) \eta^q(v')^\top} O K^\top(v-v').
\end{align}
Eq.~\eqref{eq:master_flow} and~\eqref{eq:corr_noise_def} imply that $\E{A^q(u) \eta^q(v')^\top} = \frac{1}{\tau}G(u-v')\diag{(Y)}$, so the previous expression becomes
\begin{align}
    & \E{\Delta A^q(t) \Delta q^\top(t) } \nonumber\\
    &= \frac{1}{\tau}\iint_{t-\Delta t}^t {\rm d}u{\rm d}v \int_{-\infty}^v {\rm d}v'G(u-v') \diag{(Y)}OK^\top(v-v').
\end{align}
 For $\tau\ll 1$, we have $\frac{1}{\tau} G(t-t') \xrightarrow[]{} R\delta(t-t')$ \citep{LeCozBouchaud-2024a}. It yields
\begin{align}
    \E{\Delta A^q(t) \Delta q(t) } &= R  \diag{(Y)}O\iint_{t-\Delta t}^t {\rm d}u {\rm d}vK^\top(v-u)\nonumber \\
    &= \Delta t R \diag(Y) O \int_{0}^{\Delta t} {\rm d}\ell K^{\top} (\ell).
\end{align}
Thus, the covariance between the forward rates and the order flows reads
\begin{align}
    \label{eq:covariance}
    \E{\Delta f(t)\Delta q^{\top}(t) } \hspace{3cm} \nonumber \\
    = \Delta t \diag{\sigma} \diag(\sigma_A)^{-1}R\diag(Y) O \int_{0}^{\Delta t} {\rm d} \ell K^{\top}(\ell),
\end{align}
having noted that $\E{\Delta A^{\perp}(t) \Delta q^{\top}(t)} = 0$.

\subsection{Computation of the response matrix}
We define the cross-impact matrix $\Lambda \in \mathbf{M}_n(\mathbb{R})$ as the matrix ensuring a linear relationship between forward rates increments and order flows, i.e., 
\begin{align}
    \label{eq:definition_lambda_ap}
    \Delta f(t) = \Lambda \Delta q(t) + \mathcal{E}(t),
\end{align}
where $\mathcal{E}$ is a temporally uncorrelated noise independent from $\Delta q$. One can reformulate~\eqref{eq:definition_lambda_ap} as
\begin{align}
    \label{eq:lambda_product}
    \E{\Delta f(t)  \Delta q(t)^\top  } = \Lambda \E{\Delta q(t) {\Delta q(t)}^\top  }.
\end{align}
Yet, the variance covariance matrix of the daily trading flow imbalance reads
\begin{align}
    & \E{\Delta q(t) \Delta q^\top(t)} = \iint_{t-\Delta t}^{t} {\rm d}s{\rm d}s'   \E{ \frac{{\rm d}q}{{\rm d}t}(s) \frac{{\rm d}q}{{\rm d}t}^\top (s') } \nonumber \\
    = & \iint_{t-\Delta t}^{t} {\rm d}s{\rm d}s'  \int_{-\infty}^{\min(s,s')} {\rm d}u K(s-u) {K}^{\top}(s'-u) \nonumber \\
    = & \Delta t\int_{0}^{\Delta t} {\rm d}l \int_{\mathbb{R}} {\rm d}u K(u) {K}^{\top}(u-l).
\end{align}
 Hence, replacing the left-hand side in~\eqref{eq:lambda_product} by its expression in~\eqref{eq:covariance}, we have
\begin{align}
    \label{eq:formula_lambda}
    \diag{\sigma} \diag(\sigma_A)^{-1} R \diag(Y) O \int_{0}^{\Delta t} {\rm d} l K^{\top}(l) = \nonumber \\ \Lambda \int_{0}^{\Delta t} {\rm d}l K \ast K^{\top}(l).
\end{align}
In the most general case, Eq.~\eqref{eq:formula_lambda} requires $\Lambda$ to be time-dependent. Yet, assuming ${\rm d}q$ has no temporal correlation i.e., $K(u) = \delta(u)\Omega^{1/2}$, Eq.~\eqref{eq:formula_lambda} reads
\begin{align}
    \Lambda = \diag{\sigma} \diag(\sigma_A)^{-1}R\diag(Y) O \Omega^{-1/2},
\end{align}
where the equal-time variance-covariance matrix $\Omega$ is defined by $\Omega := \frac{\E{\Delta q(t) \Delta q^{\top}(t) }}{\Delta t}$.

\section{Correlation between forward rates and flows} \label{Correlation between forward rates and flows}
In this section, we demonstrate that the correlation between the forward rate of tenor~$\theta$ and the martingale component of the order flow of tenor~$\theta'$, given by,
\begin{align}
    \label{eq:corr_price_volume}
    & \rho(\Delta f_{\theta}(t),\Delta \tilde{q}_{\theta'}(t)) = \sum_{\theta''=1}^{n} \frac{R_{\theta\theta''}}{(\sigma_A)_\theta} Y_{\theta''} \frac{\left(\Omega^{1/2}O^\top \right)_{\theta'\theta''}}{\sqrt{\Omega_{\theta'\theta'}}},
\end{align}
is well-defined.
We assume $Y=1$, the unit vector. In the case $\psi \ll 1$, $(\sigma_A)_\theta = \sqrt{\left(\mathcal{M}^{-1}\mathcal{J}^2(\mathcal{M}^{-1})^\top\right)_{\theta\theta}}$ and $R_{\theta\theta'}=\left(\mathcal{M}^{-1}\mathcal{J}\right)_{\theta\theta'}$. Equation~\eqref{eq:corr_price_volume} is then the canonical inner product of two normalized vectors of $\mathbb{R}^n$. Thus, the correlation $\rho(\Delta f_{\theta}(t),\Delta \tilde{q}_{\theta'}(t))$ is well defined. 

In the case $\psi \gg 1$, one can express the numerator and the denominator in Eq.~\eqref{eq:corr_price_volume} using the inner product~\eqref{eq_inner_product} on the space of integrable real-valued functions on $[0,\pi]$. For $(\theta,\theta') \in \llbracket1,n\rrbracket^2$, having noted that $(\sqrt{2}\cos{\xi \theta},\sqrt{2}\cos{\xi \theta'})=\delta_{\theta\theta'}$, we have
\begin{align}
    &\sum_{\theta''=1}^{n} (R)_{\theta\theta''} \left(\Omega^{1/2}O\right)_{\theta'\theta''} = \sum_{\theta''=1}^{n} (\mathcal{D}_1)_{\theta\theta''} \left(\Omega^{1/2}O\right)_{\theta'\theta''} \nonumber \\ 
    &=\left( \frac{\sqrt{2}\cos{\xi \theta}}{L_{d}(\xi)}, \sqrt{2} \sum_{\theta''=1}^{n} \cos{\xi \theta''}\left(\Omega^{1/2}O\right)_{\theta'\theta''} \right),
\end{align}
and,
\begin{align}
    & (\sigma_A)_\theta \Omega_{\theta'\theta'} = (\mathcal{D}_2)_{\theta\theta} \Omega_{\theta'\theta'} \nonumber \\
    & \hspace{1cm} = \left(\frac{\sqrt{2}\cos{\xi \theta}}{L_{d}(\xi)},\frac{\sqrt{2}\cos{\xi \theta}}{L_{d}(\xi)}\right) \times \\ \nonumber
    & \left( \sqrt{2} \sum_{\theta''=1}^{n} \cos{\xi \theta''} \left(\Omega^{1/2}O\right)_{\theta'\theta''} , \sqrt{2} \sum_{\theta''=1}^{n} \cos{\xi \theta''} \left(\Omega^{1/2}O\right)_{\theta'\theta''} \right).
\end{align}
Thus, Cauchy-Schwarz's inequality ensures 
\begin{align}
   -1\leq \frac{\sum_{\theta''=1}^{n} (\mathcal{D}_1)_{\theta\theta''} \left(\Omega^{1/2}O\right)_{\theta'\theta''} }{\sqrt{(\mathcal{D}_2)_{\theta\theta} \Omega_{\theta'\theta'}}} \leq 1.
\end{align}
\end{document}